\documentclass[prx,twocolumn,showpacs,longbibliography,nofootinbib, superscriptaddress]{revtex4-1}
\usepackage{graphics}
\usepackage{graphicx}
\usepackage{amssymb}
\usepackage{epstopdf}
\usepackage{color}
\usepackage{hyperref}
\usepackage{url}
\usepackage[normalem]{ulem}
\usepackage{mhchem}

\DeclareGraphicsExtensions{.pdf, .png, .jpg, .eps}
\newcommand{\bfig}{\begin{figure}}
\newcommand{\efig}{\end{figure}}

\usepackage{float}

\newcommand{\bea}{\begin{eqnarray}}
\newcommand{\ena}{\end{eqnarray}}
\newcommand{\bee}{\begin{equation}}
\newcommand{\ene}{\end{equation}}

\newcommand{\black}{\color{black}}

\newcommand{\cers} {CsErSe$_{2}$ }
\newcommand{\erion} {Er$^{3+}$}

\newcommand{\hcef} {\mathcal H_{\rm CEF}}
\newcommand{\hxxz} {\mathcal H_{\rm xxz}}
\newcommand{\hz} {\mathcal H_{\rm Z}}
\newcommand{\htot} {\mathcal H_{\rm tot}}

\newcommand{\Jxx} {\mathcal J_{\perp}}
\newcommand{\Jzz} {\mathcal J_z}

\newcommand{\Bpb} { \mathbf B\parallel b}
\newcommand{\Bpc} { \mathbf B\parallel c}
\newcommand{\Bnm} {B_n^m}
\newcommand{\Bc} {B_{\mathrm C}}
\newcommand{\Bx} {B_{\mathrm x}}

\newcommand{\Onm} {\hat O_n^m}

\newcommand{\invcm} { cm$^{-1}$}
\newcommand{\chiac} { \chi^{\rm ac}_c}

\renewcommand{\arraystretch}{1.25} 

\begin{document}

\title {
Field-induced magnetic phase transition driven by a  ground state level-crossing in CsErSe$_2$ \black }

\author{Hope Whitelock}
\affiliation{Department of Physics, University of Colorado, Boulder, Colorado 
80309, USA}

\author{Allen O. Scheie}
\affiliation{Los Alamos National Laboratory, Los Alamos, New Mexico, USA}

\author{Marissa McMaster}
\affiliation{Department of Physics, University of Colorado, Boulder, Colorado 
80309, USA}

\author{Ian A. Leahy}
\affiliation{Department of Physics, University of Colorado, Boulder, Colorado 
80309, USA}

\author{Li Xiang }
\affiliation{National High Magnetic Field Laboratory, Tallahassee, Florida, USA}

\author{Mykhaylo Ozerov}
\affiliation{National High Magnetic Field Laboratory, Tallahassee, Florida, USA}

\author{Dmitry Smirnov}
\affiliation{National High Magnetic Field Laboratory, Tallahassee, Florida, USA}

\author{ Eun Sang Choi }
\affiliation{National High Magnetic Field Laboratory, Tallahassee, Florida, USA}

\author{C. dela Cruz}
\affiliation{Neutron Scattering Division, Oak Ridge National Laboratory, Oak Ridge, Tennessee 37831, USA}

\author{M. O. Ajeesh}
\affiliation{Los Alamos National Laboratory, Los Alamos, New Mexico, USA}
\affiliation{Department of Physics, Indian Institute of Technology Palakkad, Kerala 678623, India}

\author{Eliana S. Krakovsky}
\affiliation{Department of Materials Science \& Engineering, Stanford University, Stanford, CA 94305, USA}
\affiliation{Computational Physics Division, Los Alamos National Laboratory, Los Alamos, New Mexico 87545, USA}

\author{Daniel A. Rehn}
\affiliation{Computational Physics Division, Los Alamos National Laboratory, Los Alamos, New Mexico 87545, USA}

\author{Jie Xing}
\affiliation{Neutron Scattering Division, Oak Ridge National 
Laboratory, Oak Ridge, Tennessee 37831, USA}

\author{Athena S. Sefat}
\affiliation{Materials Science and Technology Division, Oak Ridge National 
Laboratory, Oak Ridge, Tennessee 37831, USA}

\author{Minhyea Lee}
\email{minhyea.lee@colorado.edu}
\affiliation{Department of Physics, University of Colorado, Boulder, Colorado 
80309, USA}

\date{\today}

\begin{abstract}
We report a comprehensive study of the low-temperature magnetic properties of the insulating rare-earth triangular magnet CsErSe$_2$.
We uncover a field-induced level crossing at the crystal electric field (CEF) ground state, which gives rise to a first order phase transition as well as to distinctive magnetic properties. 
This crossing is identified by the accurate determination of the single-ion Hamiltonian, the reliability of which is substantially enhanced by field-dependent optical spectroscopy that directly tracks the Zeeman splittings of Kramers doublets.
We also observe spontaneous antiferromagnetic ordering in CsErSe$_2$ below $T_N \approx 110$ mK, and resolve the corresponding magnetic structure using elastic neutron scattering. 
We discuss how the rich magnetic behavior of \cers arises from the interplay of non-trivial field-dependent single-ion physics and spontaneous ordering, and highlight the implications of these results for understanding  magnetic phenomena across a wide range of insulating rare-earth magnets. 
\end{abstract}

\maketitle

\section{Introduction}
\label{intro}

A primary goal at the interface of theoretical and experimental efforts in quantum magnetism is to understand exotic spin states, most notably quantum spin liquids, which host fractionalized excitations, long-ranged entanglement, and topological order \cite{SavaryReview2016, JWen2019}. 
Magnetic insulators with strong spin-orbit coupling (SOC) are now widely recognized as a promising platform for accessing such unconventional ground states by extending the nature of magnetic frustration. 

Compounds based on $4f$ rare-earth (RE) ions  have provided  fertile  ground  for quantum magnetism research, thanks to their unique combination of geometric frustration and strong SOCs \cite{Gardner2010, JRau2019, Smith2025}. 
In these materials, however, a detailed understanding of the microscopic physics is a prerequisite for developing minimal spin models for effective theories of many-body phenomena \cite{Gardner2010, JRau2019}.

The key to a microscopic description of the magnetic interactions in RE magnets is a reliable determination of the single-ion crystal electric-field (CEF) Hamiltonian ($\hcef$), comprising of the linear combination of  Stevens' operators ($\Onm$'s) allowed by the site symmetry of RE ion and their coefficients ($\Bnm$'s) CEF effects split the degeneracy of the ground multiplet of the total angular momentum $J$ \cite{Stevens1952, Hutchings1964}. 
This produces a set of Kramers doublets for  RE ions with half-odd-integer $J$. 
At sufficiently low temperatures, the resulting dynamics are confined to the lowest doublet, thereby justifying an effective pseudospin-1/2 description and enabling the application of the extensive theoretical framework developed for frustrated magnetism. 

\begin{figure}[ht]
    \centering
    \includegraphics[width=\linewidth]{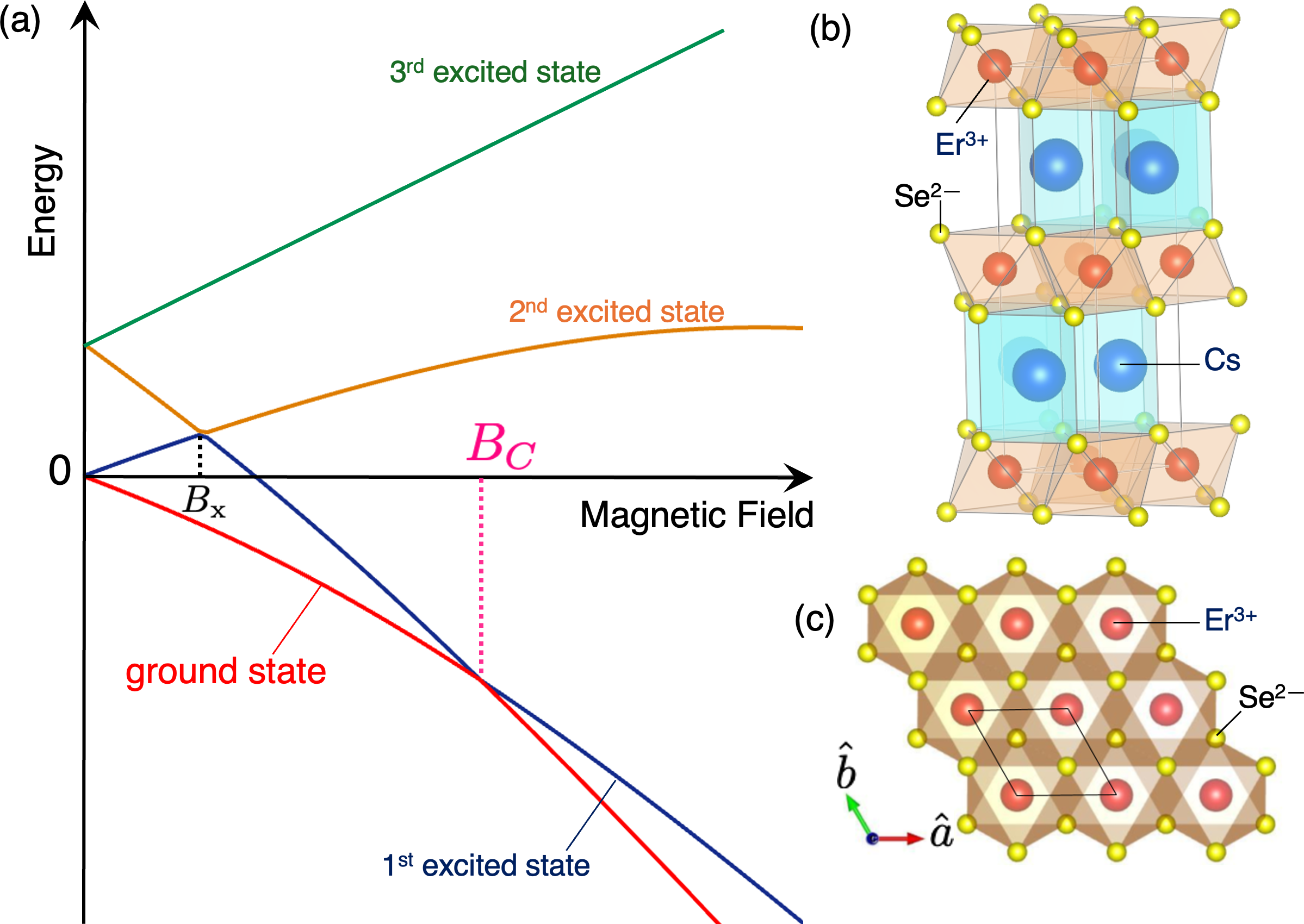}
    \caption{(a) Schematic sketch  for field-induced level splitting of the two lowest energy doublets in \cers is shown. The ground state undergoes a first-order quantum phase transition at  $\Bc$, where the derivative of the energy with respect to $B$  discontinuously changes. Another crossing between the first and second excited states shown at $\Bx$. 
    (b) Crystal structure of  \cers adopting $P6_3/mmc$ space group \cite{Xing2020acs}. (c)  \erion (red) triangular layers on the $ab$-plane formed by edge-sharing ErSe$_6$ octahedra, exhibiting $D_{3d}$ site symmetry
 }
    \label{fig:sch}
\end{figure}

The delafossite ($A$RESe$_2$, $A=$ K, Na, Cs, RE = rare-earth ions) family, which hosts a 2-dimensional triangular net of RE ions layered along the $c$-axis, \cite{Xing2020acs} has attracted much attention. Of particular interest are the compounds with Kramers' ions such as Yb$^{+3}$ ($J=7/2$)\cite{Bordelon2019, PDDai2021, Zhang2021,Pocs2021, Scheie2024} and Er$^{+3}$ ($J=15/2$)~\cite{Scheie2020Er, Xing2019prm,GDing2023}, which undergo Zeeman splitting in magnetic fields~\cite{Pocs2021}. 
When multiple CEF levels are closely spaced,  their field-induced splittings can host level crossings, resulting in distinctive field-dependent spectra. 
When such a crossing occurs in the ground state, it gives rise to a field-induced phase transition, profoundly impacting the magnetic properties. 
Fig. \ref{fig:sch} (a) schematically shows the Zeeman splittings of the two lowest-lying doublets upon applying a magnetic field.  
The discontinuous change in the slope of the ground state 
energy \black at the critical field $B=\Bc$ indicates a first order phase transition at zero temperature.  
Furthermore, the crossing of excited states, marked as $\Bx$, results in the non-monotonic Zeeman gap that closes and reopens at $\Bc$, which has a profound impact to give a rise to complex and rich field dependence of magnetic properties. 
 
In this work, we identify a field-induced phase transition arising from the level-crossing in the ground state of the delafossite  CeErSe$_2$ at $\Bc \approx 5.4$ T, when magnetic field ($B$) is applied along the crystalline $c$-axis ($\Bpc$). 
To our best knowledge, such field-induced level-crossings involving the ground state and the corresponding field-induced transition have not been explicitly predicted from a well-characterized CEF Hamiltonian. \black

This first-order transition is manifested in a step-like change in magnetization ($M_c$) and a corresponding conspicuous peak at $\Bc$ in  magnetic susceptibility as a function of field. 
We performed AC susceptibility measurements and verified an peak at $\Bc$. \black
In $B<\Bc$,  the non-monotonic Zeeman gap exhibits its maximum at  $\Bx$ --- the crossing of the 1st and 2nd excited states (Fig. \ref{fig:sch}(a)) --- which leads to a plateau-like field dependence of magnetization before its step-like increase at $\Bc$. 

We predicted this transition and associated magnetic properties based one our model Hamiltonian $\htot$ for the system, which consists of the CEF  ($\hcef$), Zeeman ($\hz$), and XXZ-like exchange interactions ($\hxxz$) within the Weiss mean-field (MF)  approximation:  
\bee
\htot = \hcef+ \hz+\hxxz. 
\label{eq:htot}
\ene
Because of a weak contribution from exchange interaction in the $f$-electron systems like \cers,  the robust determination of $\hcef$ is most crucial to accurately predicting magnetic behavior.   
We achieve this by directly probing the field-dependence of the CEF level splitting using magneto-optical spectroscopy. 
We find  that leveraging the magnetic field offers a unique opportunity to determine the parameters of $\hcef$, leading us to accurate predictions.

Finally, we report long-range magnetic ordering in \cers below $T_N = 110$ mK. 
Our neutron diffraction reveals stripe antiferromagnetic ordering.   
Experimentally observed $T_N$ is consistent with in-plane exchange energy obtained from our fitting, which reassures us that the validity of the MF approximation extends to low temperatures.  \black

With an accurately determined microscopic Hamiltonian,  our work reveals the rich CEF physics of a single ion under high magnetic field. 
This opens a new avenue for exploring frustrated magnetism in rare-earth systems, enables identification of the origin of unusual magnetic properties, and establishes a firm foundation for disentangling collective inter-ion interactions from single-ion physics.

This paper is organized as follows. Sec. \ref{methods} introduces the
CsErSe$_2$ sample and experimental methods.  
In Sec. \ref{MO}, we first present magneto-optical spectroscopic data obtained from far-infrared and Raman experiments and the determination of the set of parameters in the Hamiltonian in Eq. (\ref{eq:htot}) based on those data.  
Sec. \ref{chiac} reports AC magnetic susceptibility measurements as a function of temperature and field, compared with the predictions of our Hamiltonian.  
Sec. \ref{neutron} presents the long-range magnetic order revealed by elastic neutron
scattering experiments. 
Sec. \ref{disc} discusses the overall implications of our work, and Sec. V summarizes the study.

\section{Methods}
\label{methods}
\cers crystallizes in a $P6_3/mmc$  space group  \cite{Xing2020acs, Scheie2020Er} as illustrated in Fig.~\ref{fig:sch}(b). 
The triangular-lattice planes [Fig. \ref{fig:sch}(c)] consist of edge-sharing ErSe$_6$ octahedra, allowing $D_{3d}$ site symmetry for \erion, and stacks along the $c$-axis. 
Within the $ab$-plane, the Se atoms mediate antiferromagnetic (AFM) superexchange interactions between nearest-neighbor \erion ion pairs.

Single crystals of \cers used in these experiments were grown from powder samples using CsCl flux as described in \cite{Xing2020acs}. 
Typical sizes  of single crystals  used for optical spectroscopy are $(1-2)\times (0.3- 0.8) \times (0.02-0.04)$ mm$^3$. 

The magneto-optical spectroscopy was performed with both Raman and far-infrared (FIR) measurements at the National High Magnetic Field Laboratory.  
Raman experiments were done in a Quantum Design Physical Property Measurement System with a magnetic field up to 14 T. 
Scattered light was collected in backscattering geometry from the $ab$ plane of a  \cers crystal with the magnetic field applied along the $c$axis (Faraday geometry) using an unpolarized 532 nm laser focused to a $1-2 ~\mu$m-diameter spot with the incident power of  0.5 -- 1 mW. 
Collected signal spectra were analyzed using a monochromator (HRS750, Teledyne Princeton Instruments, 1200 g/mm grating) and recorded by a liquid-nitrogen-cooled CCD (PyLoN:400BR, Teledyne Princeton Instruments) with a spectral resolution of approximately 1.3 \invcm.

FIR spectroscopy was performed in a 17.5 T superconducting magnet with a Fourier-transform infrared 
spectrometer 
by  Bruker Vertex 80v. 
Broadband IR light provided by the FTIR spectrometer is guided via evacuated brass light pipes to the sample space, situated at the center of the magnet.
Transmitted IR light is delivered to a composite Si-bolometer (IR Labs, Inc.) mounted in the same cryogenic environment, cooled by low-pressure helium gas to a temperature of approximately 5 K.  
We used both Faraday ($\Bpc$) and Voigt ($\Bpb$) geometries with the wave vector of the incoming IR light to measure the\cers  transmission spectra.
To highlight the field-dependent absorption modes, we normalized the spectrum at each field to the maximum intensity value and then subtracted the zero-field spectrum. 
The resulting normalized transmission is sensitive only to 
intensity changes induced by the magnetic field 
while eliminating contributions from field-independent vibrational absorption and instrumental artifacts.
\black

AC magnetic susceptibility was measured in a nesting coil set, consisting of a primary coil that generates a very small AC magnetic field and a pair of sensing coils, coaxially placed within the primary coil. 
The two sensing coils are counter-wound and connected in series to ensure equal mutual inductance with opposite sign. 
This results in zero net induced voltage in a null state. Six single crystal pieces of \cers samples were stacked together along the $c$-axis and placed in one of the sensing coils. 
The change of magnetic flux contributed by the sample induces a directly proportional nonzero net voltage. 
We find the background signal of coils is negligible compared to the signal generated by the presence of the samples.  
The coil set is mounted on a dilution refrigerator such that the DC magnetic field generated by the superconducting magnet is aligned to the $c$-axis.

Neutron diffraction spectra were measured at the HB2A 
diffractometer \black at the Oak Ridge National Laboratory's High Flux Isotope Reactor \cite{garlea2010high}.
We measured 3 g of loose powder of \ce{CsErSe2} (the same samples as used in Ref. \cite{Scheie2020Er}) in a copper can mounted in a dilution refrigerator at temperatures $T = 0.5$ K and $T=0.05$ K using neutrons with $\lambda = 2.41$ \AA. 
The magnetic scattering was isolated by subtracting the high-temperature data from the low-temperature data. 
We performed Reitveld refinements to the neutron data using the \textit{Fullprof} software package~\cite{rodriguez1993recent}.

Phonon modes in CsErSe$_2$  are calculated from density functional theory (DFT), which is performed using the \emph{Vienna ab Initio Simulation Package} (VASP) version 5.4.4~\cite{kresse1993ab, kresse1996efficient, kresse1996efficiency, kresse1994norm} using the projector augmented wave (PAW)~\cite{blochl1994projector, kresse1999ultrasoft} method and the PBE exchange-correlation functional~\cite{perdew1996generalized}.  
A $\Gamma$-centered Monkhorst-Pack~\cite{PhysRevB.13.5188} k-mesh is used for all calculations. Phonons are calculated using the Phonopy software~\cite{phonopy-phono3py-JPCM,phonopy-phono3py-JPSJ}. \black

\section{results}

\subsection {Magneto-optical spectroscopy and  the determination of $\hcef$}
\label{MO}
\begin{figure*}
    \includegraphics[width=.95\linewidth]{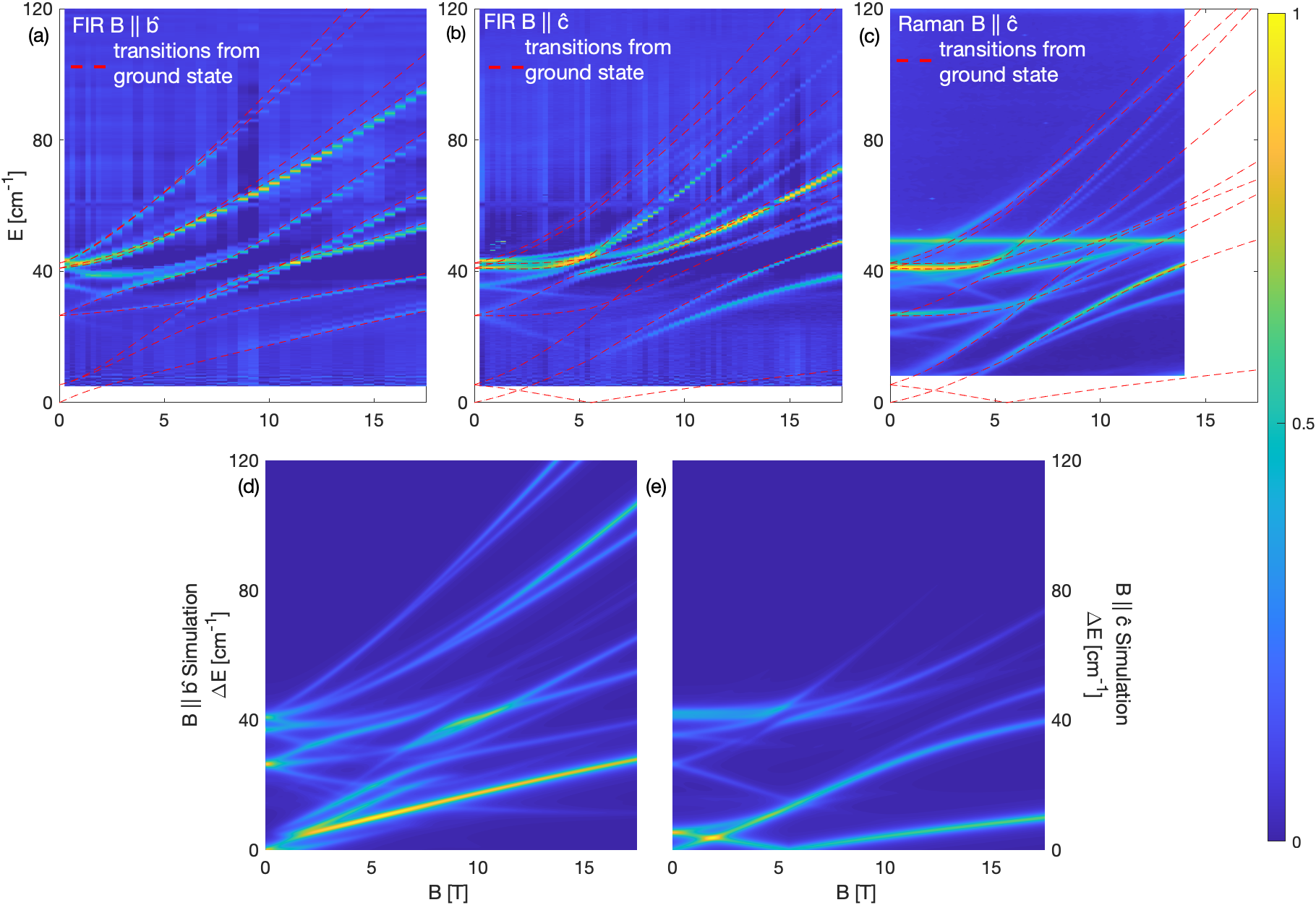}
    \caption{ Normalized FIR 
    absorption spectral intensity \black at $T=5$ K  
    as functions of incoming light energy  $E$ and the applied magnetic field $B$ \black for the two high-symmetry field orientations,  $\Bpb$ (a) and $\Bpc$ (b), respectively. (c) Normalized peak intensity of Raman shift, denoted as $E$  in $\Bpc$ at $T = 5$ K.  
    Red dashed lines in (a-c) show the field-dependent transition from the CEF ground state as a function of $B$ calculated from our analysis for $E<120$ \invcm (See text). \black
  (d-e) 
  Simulated transition intensity spectrum \black for $E<120$ \invcm for (d) $\Bpb$ and  (e) $\Bpc$.
  }
    \label{fig:spec}
\end{figure*}

The FIR and Raman spectra observed under a magnetic field are shown in Fig.  \ref{fig:spec} (a-c):  the panels (a) and (b) display colormaps of the field dependence of the normalized FIR absorption measured at $T=5$ K, with applied field ($B$) up to 17.5 T along the crystalline $b$ axis ($\Bpb$) and $c$-axis, respectively. 
Panel (c) displays the Raman shift at  5 K as well, as a function of field up to 14 T along the $c$-axis ($\Bpc$), the intensity of which is normalized with the maximum value.  
The red 
dashed \black lines in each figure indicate the calculated peak positions of the transitions from the ground state.  
Strong single-ion magnetoanisotropy of \cers for the different field orientations is self-evident from the distinctly anisotropic field dependence of the CEF levels. 

We note that the FIR spectra of \cers~suffer from complete absorption in  $130\leq E \leq 210$ \invcm, and as such, above  210 \invcm,  few discernible field-dependent features were observed.  
The Raman shift data display three field-dependent spectra in $ E > 120$ \invcm [Fig. \ref{fig:raman}], however, with far weaker intensity. 
It turns out that three highest-energy doublets exhibit an exceptionally small transition probability  that cannot be discernible. We will discuss further in Sec.\ref{disc}.
Hence, our analysis is focused on only field-dependent spectra originating from CEF levels below $E< 120$ \invcm.   
\black

Despite the fundamentally different nature of FIR and Raman in $\Bpc$, the close correspondence between panels (b) and (c) is remarkable. 
Apart from the field-independent peak at $E \approx 49$ \invcm in panel (c) the two techniques clearly probe the same transitions between CEF energy levels. 
\black
The field-independent line at 49 \invcm  is identified as Raman-active $E^1_ {2g}$ phonon from the phonon bands calculation as shown in Table \ref{tb:phonon}. 
%

\begin{figure}[ht]
    \includegraphics[width=\linewidth]{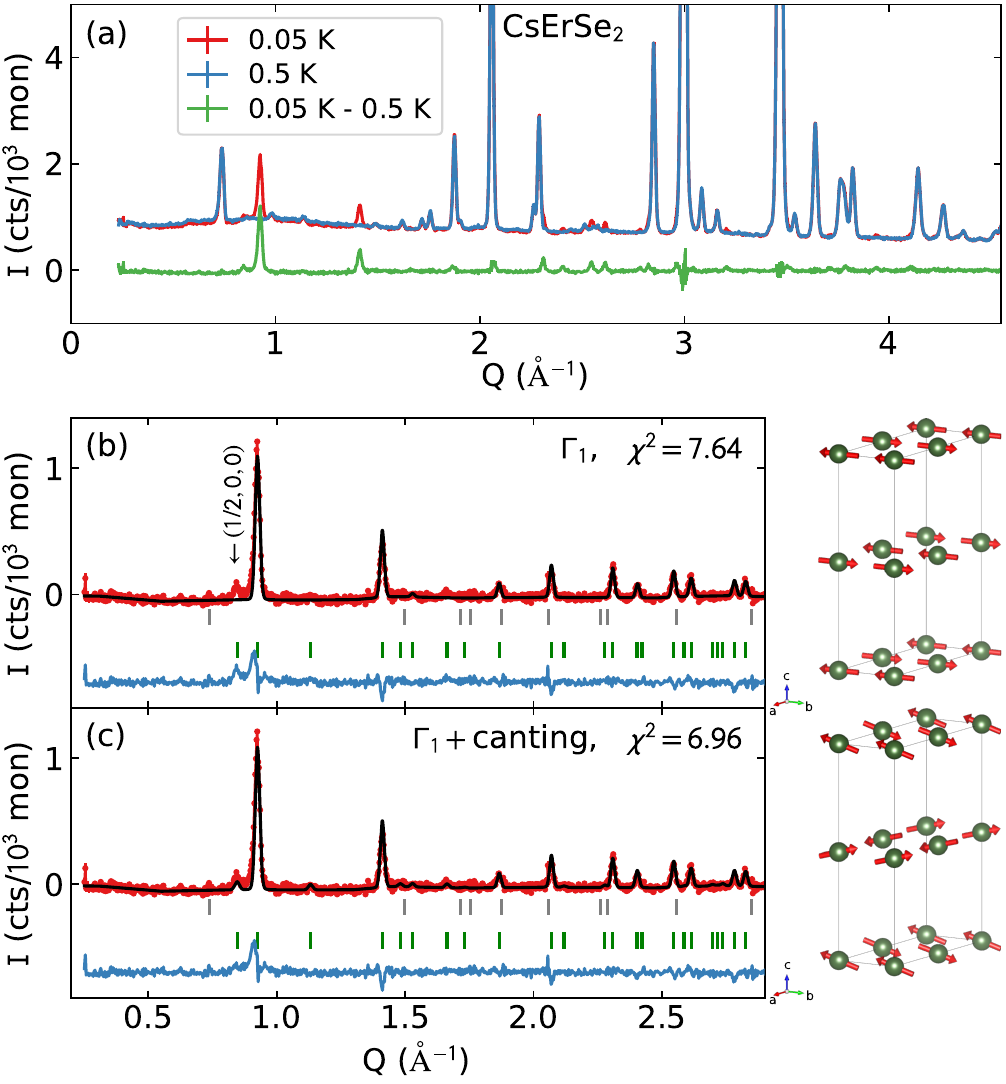}
    \caption{ (a) Neutron diffraction on CErSe$_2$, showing magnetic reflection with propagation vector 
    $\mathbf k = (\frac{1}{2},0,0)$ appearing at $T = 50$ mK which is about 1/2 of T$_N \approx 110$ mK. \black (b-c) Refinements of the \ce{CsErSE2 magnetic structure with }
    $(\frac{1}{2},0,1) \Gamma_1$ order and with an added $(\frac{1}{2},0,0)$ canting along the $c$-axis. This additional canting accounts for the $(\frac{1}{2},0,0)$ Bragg peak, and noticeably lowers the reduced $\chi^2$. The refined moment is 71\% of the theoretical maximum value from the CEF ground state doublet 3.27(15)$\mu_B$, indicating weak fluctuations.}
    \label{fig:neutron}
\end{figure}

FIR and Raman shift measurements reveal the transitions from the lowest states to the Zeeman split excited states of the \erion~ion. \black
In order to describe the intricate field dependencies shown above, we consider the three contributions in our Hamiltonian as expressed in Eq. (\ref{eq:htot}).  
$\hcef$ describes the single-ion energy scales  in terms of  the linear combination of the six Stevens operators allowed by \erion's site symmetry,
$\hcef = B_2^0\hat{O}_2^0 + B_4^0\hat{O}_4^0
 + B_4^3\hat{O}_4^3 + B_6^0\hat{O}_6^0 + B_6^3\hat{O}_6^3
 + B_6^6\hat{O}_6^6$. 
This plays the most crucial role in determining the energy landscape of \cers: at zero field, the energy eigenstates of $\hcef$ split the $J = 15/2$ multiplet into eight Kramers doublets. 
Upon applying  field, each doublet splits into two states according to the Zeeman interaction, which corresponds to the second term in Eq. (\ref{eq:htot}): $\hz = -  \mu_B g_J \mathbf{B} \cdot \sum_{i} \hat{\mathbf{J}},$ where $\hat{\mathbf{J}}$ refers to the total angular momentum operators and Land\'e $g$-factor, $g_J = 6/5$, is used for \erion.
%
Beyond the single ion picture, the triangular lattice symmetries of Er$^{3+}$ sites permit a nearest-neighbor exchange interaction with XXZ symmetry \cite{Yamamoto2014}. Since the pseudodipolar anisotropic exchange terms vanish within a Weiss mean-field treatment,\cite{Maksimov2019} we confine our analysis to the minimal XXZ  $\hxxz$, featuring  nearest-neighbor $J=15/2$ moments coupled by $\Jxx$ and $\Jzz$ for transverse  and longitudinal $J$ components: 
$\hxxz = \sum_{\langle i,j\rangle}[
\Jxx ( \hat{J}_{i,x} \hat{J}_{j,x} + \hat{J}_{i,y} \hat{J}
_{j,y})  + \Jzz \hat{J}_{i,z} \hat{J}_{j,z}]$, where the indices $\langle i,j \rangle$ refer to nearest-neighbor 
lattice sites and $x,y,z$, label 
the components of  $J = 15/2$ moment on site $i$ \cite{note1}. 
 In addition, there is a nonzero $c$-axis exchange which produces the long-range order shown in Fig. \ref{fig:neutron}, which is not included here.\black

\begin{table}[b!]
    \centering
    \setlength{\tabcolsep}{8pt}
    \renewcommand{\arraystretch}{1.25}
    \begin{tabular}{cc|cc}
        \hline \hline
         &    meV   & & meV \\
        \hline
        $B_2^0$ & $-3.721(1) \times 10^{-2}$ &  $B_4^0$ &$-3.880(1) \times 10^{-4}$\\
        $B_4^3 $&$-1.407(1) \times 10^{-2}$ &$B_6^0$ &$+3.187(1) \times 10^{-6}$  \\
        $B_6^3$ &$ -3.593(1) \times 10^{-6}$ &  $B_6^6$ &$+3.491(1) \times 10^{-5}$  \\
        \hline\hline
        $\mathcal J_z$ & $-2.64(1) \times 10^{-3}$ & $\mathcal J_{\perp}$  & $-0.53(1) \times 10^{-3}$ \\
          \hline \hline
    \end{tabular}
    \caption{\small  The values of the coefficients of Stevens operators of \ce{CsErSe2} obtained from fits to Raman shift in Fig.~ \ref{fig:spec} (c). Units of all shown quantities are meV.}
    \label{tb:bnmval}
\end{table}

To specify $\htot$ in Eq. (\ref{eq:htot}), a total of eight parameters -- six coefficients of Stevens' operators, $\Bnm$'s in $\hcef$, and two exchange energies in $\hxxz$ --  are required. 
At a given high-symmetry field orientation, either $\Bpc$ or $\Bpb$, there are seven free parameters that must be determined.   
%
 We fit the field-dependent spectra in the Raman shift data  [Fig.\ref{fig:spec} (c)] directly to  Eq. (\ref{eq:htot}), leveraging the field dependence and obtain seven parameters simultaneously for $\Bpc$.  \black
The fittings were performed using the scipy optimize package \cite{2020SciPy-NMeth}, and PyCrystalField \cite{Scheie2021py}.

The fitting parameters obtained are listed in Table \ref{tb:bnmval}. They agree well with values obtained using an entirely independent experimental method -- zero-field inelastic neutron spectroscopy and low-field magnetization measurements \cite{Scheie2020Er}. This agreement highlights the accuracy of our CEF parameter determination, which is essential for predicting magnetic properties governed by single-ion physics in rare-earth compounds.
\black


Fig. \ref{fig:spec} (d) and (e) display the calculated normalized transition amplitudes for $\Bpb$ and $\Bpc$, respectively. 
They are calculated by Fermi's golden rules using the results of solving the eigen equation for  Eq.(\ref{eq:htot}) with the parameters in Table \ref{tb:bnmval}.  
The majority of the field-dependent spectral lines observed experimentally in Fig. \ref{fig:spec} (a-c) are captured in Fig.\ref{fig:spec}(d) and (e).

In particular, the red lines in panels (a-c) represent the peak positions of transitions from only the ground state to higher excited states, calculated from our model Hamiltonian. In $\Bpc$ geometry, the calculations reveal multiple level crossings, including one involving the ground state. However, the direct experimental observation of the crossing on the  ground state is precluded by the low energy cutoff of our spectroscopic techniques.  \black 

%
The energy gap between the two lowest doublets at zero field is approximately 5.3 \invcm ( $\approx 0.67$ meV), corresponding to a temperature scale about 8 K. Thus, at the nominal temperature of the measurement $T=5$ K,  there can be thermal populations above the ground state, which allows finite transition amplitudes to higher-energy states [Fig. \ref{fig:sch}(a)]. We will discuss this further in Sec. \ref{disc}.
We also note that, as seen in Table \ref{tb:bnmval}, the exchange energies in \cers~ are on the order of $10^{-3}$ meV or less, far smaller than both the CEF interaction and the Zeeman splitting, typical for $f$-electron systems.  Consequently, the observed magnetic properties, including conspicuous magnetic anisotropy, are governed by the single-ion ground state properties.


\subsection{Field dependence of  AC magnetic susceptibility}
\label {chiac}

The most striking feature revealed by our model Hamiltonian is the ground state inversion at $\Bc \approx 5$ T [Fig. \ref{fig:spec} (c) and (e)], indicating a field-induced first order phase transition.
To examine the corresponding macroscopic magnetic properties, we present the AC magnetic susceptibility ($\chiac$) measurements as a function of temperature and field in $\Bpc$.

\begin{figure}[t]
    \includegraphics[width=\linewidth]{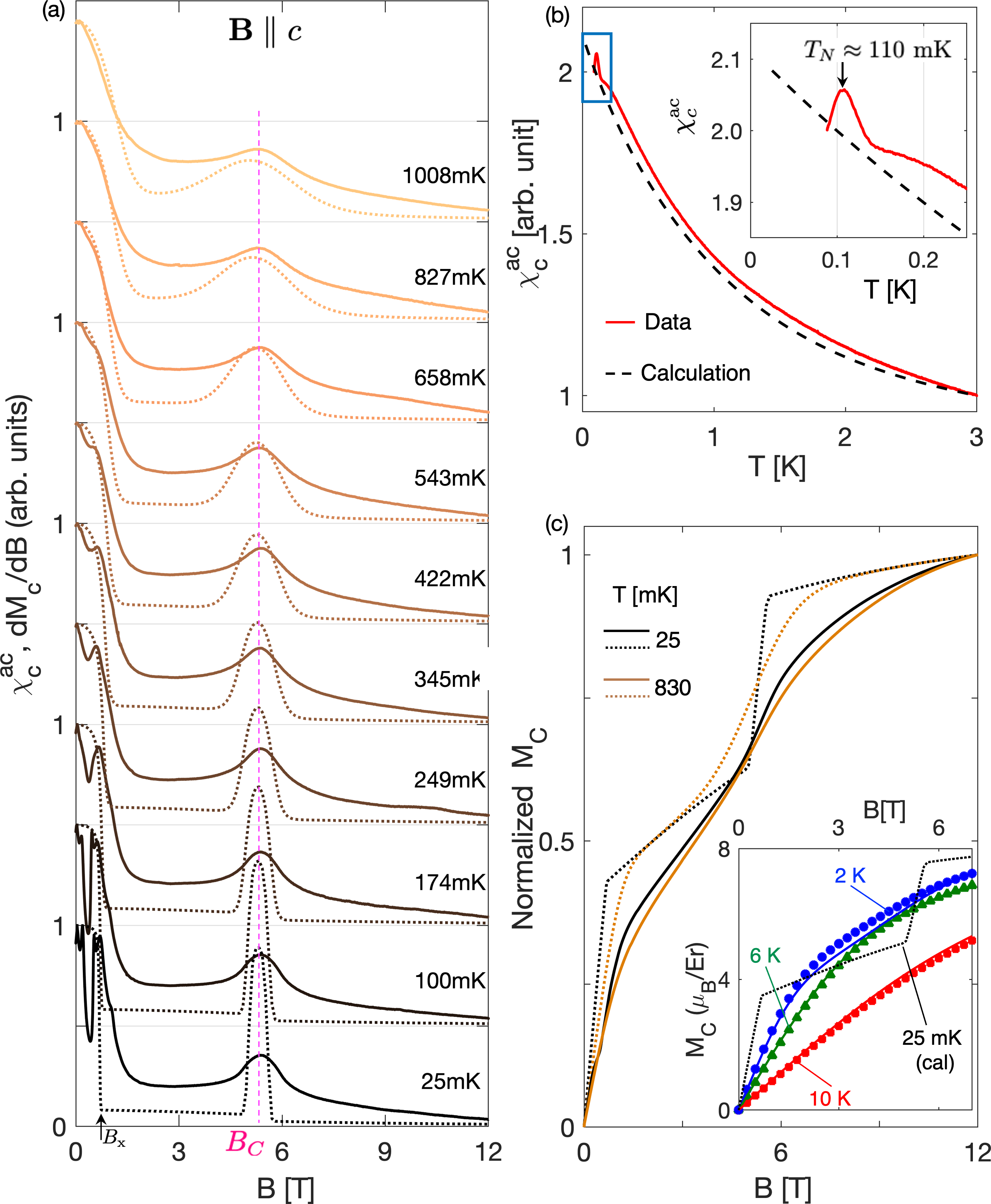}
    \caption{(a) Measured  
    susceptibility, $\chiac$, \black are shown as a function of field at various temperatures in solid lines. Dotted lines display calculated $\chiac$. Offset added for clarity. The vertical broken line indicates $B_C\approx 5.4$, where the first excited state crosses the ground state, while a black  arrow marks $\Bx$, where the first and second excited states cross. 
    (b) $\chiac~(T)$ measured at $B=0.05$ T is shown as a solid red line, with the corresponding calculation shown as a black dashed line.    A peak of $\chiac$ at  $T = 110$ mK indicates a long-range order, but is not captured in the calculation. 
    (c)  Calculated $M_c (B)$ (dotted) and normalized magnetization $M_c(B)$ obtained from numerical integration of $\chiac(B)$ (solid)  at $25$ mK and 830 mK. All curves normalized at $B=12$ T for comparison.  Inset: Experimentally measured magnetization $M_c$ (symbols)  and calculated magnetization (solid lines) at 2 (blue), 6 (green), and 10 K (red) in solid lines. Calculated $M_c (B)$ at $T=25$ mK is shown as a dotted line for comparison. }
    \label{fig:mag}
\end{figure}
 
In Figs. \ref{fig:mag}(a) and (b), we show $\chiac(B)$ at various temperatures as indicated and $\chiac(T)$ measured at the zero field limit (at 0.05 T), respectively.
$\chiac(T)$  exhibits a sharp peak at  $T_N \approx 110$ mK (solid red line). 
This is evidence of long-range magnetic ordering arising from the interactions between magnetic ions and hence beyond the single-ion picture. The calculation of $\chiac(T)$ [Inset of Fig.\ref{fig:mag}(b)],  therefore cannot capture this peak.  
In fact, such long-range order has been previously reported in a sister compound with the same structure, KErSe$_2$  at $T_N \approx  200$ mK, with a similar stripe antiferromagnetic order \cite{Xing2021, GDing2023}.
\black
To probe the nature of the ordering further, we performed elastic neutron scattering at 50 mK. 
We will discuss this in the next Sec. \ref{neutron}.

$\chiac (B)$'s feature a prominent local maximum at $B_C\approx 5.4 $ T,  as well as a peak in the limit of $B\rightarrow 0$. We identify the former as a first order field-induced transition due to a level crossing, and subsequent ground state inversion. 
This transition is an intrinsic consequence of the single-ion physics governed by $\hcef$ and $\hz$ in Eq. (\ref{eq:htot}). Accordingly, the corresponding peak position remains at $\Bc$, independent of temperature, as indicated by the vertical dashed line in Fig. \ref{fig:mag} (a). Its amplitude, on the other hand,  decreases with increasing temperature as thermal population spreads over multiple CEF levels, yet  the feature remains clearly visible up to 1.008 K.

We compare this with the calculated $dM_c/dB$, shown as dotted lines using the parameters in Table \ref{tb:bnmval}, and find good agreement,  except for a clear broadening.  
We note that the calculated values of $dM_c/dB$ and magnetization were obtained using only the parameters listed in Table \ref{tb:bnmval}; no additional fitting parameters were introduced.
\black
The agreement is even more evident in Fig. \ref{fig:mag} (c), where the integration of the measured  $\chiac(B)$ at $T =$ 25 and 830 mK are plotted in comparison to calculated $M_c$. Upon further increasing temperature, the rapid increase in $M_c$ near $B_C$ disappears entirely by $T = 2$ K, which is approximately the lowest temperature accessible with a conventional magnetometer [inset of Fig.~\ref{fig:mag}(c)].
%
We think this is likely due to the stacking faults common to layered materials \cite {HBCao2016, Bette2019} and the consequent distortions of octahedral environments. 
This can  lead to a finite distribution of values of $B_C$, due to  the spread of CEF parameters. 
This broadening can also be attributed to partial degradation common to layered materials containing alkali metal layers [Fig. \ref{fig:sch}(b)] \cite{nco}.  \black

Another unusual property in $\chiac(B)$  of \cers~ is found near zero field, specifically  $0<B<\Bx$ in both data and calculation. The conspicuous peak in $\chiac(B)$ in the limit of  $B\rightarrow0$ is a natural consequence of the Weiss MF approximation \cite{Blundellbook}, however,  two peculiar $B$-dependent features overlap on this; 
One is the abrupt drop at $\Bx$ followed by remaining at a finite  $\chiac (B)$ in $\Bx<B<\Bc$.  
The other is an abrupt dip centered at around 0.2 T, that completely disappears for $T>830$ mK.  \black
  
The first feature is best explained by examining the lowest two levels under field, where the Zeeman gap exhibits non-monotonic field dependence, having a maximum at $\Bx$, the field corresponding to the level-crossing between the 1st and 2nd excited states [Fig. \ref{fig:sch}(a)]. 
Starting from zero field, the Zeeman gap initially increases with $B$, 
making thermal excitation across the gap more difficult at a fixed temperature and 
causing $dM_c/dB$ to decrease rapidly. For $B>\Bx$, however, the gap decreases again, so excitations become progressively easier, and the system continues to polarize until the gap closes at $\Bc$. As a result, $M_c$ increases smoothly with field, leading to a finite $\chiac(B)$ in the range of $ \Bx<B<\Bc$.

The second feature, while we find it experimentally robust, is $not$ captured by our model Hamiltonian, and at present its origin remains unclear.
We speculate that this feature is associated with the onset of the spontaneous long-range order. It appears only at low temperatures in $T< 500$ mK, the same temperature range in which the predominant stripe antiferromagnetic order with wavevector $(1/2,0,1)$  also vanishes [see Sec.\ref{neutron}]. 
This abrupt suppression of the magnetic susceptibility within the ordered phase may therefore potentially indicate a spin-flop-like transition at low fields~\cite{Blundellbook} or magnetic domains at play. \black
\black

\subsection{Magnetic structure analysis with neutron scattering }
\label{neutron}

The anomaly at $T = 110$ mK in  $\chiac (T)$  [Fig.~\ref{fig:mag} (b) indicates the formation of long-range order at zero field and prompts investigation of the nature of long range ordering in \cers.  
Long-range order has been reported in similar materials, such as \ce{KErSe2}, in both powder and single crystal samples. \cite{Xing2019prm,GDing2023}
This long-range order is mostly governed by interactions among pseudospins ($\mathcal H_{\rm XXZ}$), of which the energy scale is two orders of magnitude smaller than CEF energy scales [Table \ref{tb:bnmval}].
To gain insight into the many-body magnetic structure, we performed elastic neutron scattering  at $T=50$ mK.  
We refined the magnetic structure of \cers by performing irreducible representational analysis for the $P6_3/m m c$ space group and the $(1/2,0,1)$ propagation vector. The four irreducible representations and further information are available in \cite{supp}.
There are many possible combinations of irreducible representations that will account for this magnetic structure, but we here show the simplest option that preserves uniform moment size: alternating canting along the $c$-axis, shown in Fig. \ref{fig:neutron}(b). 
These intensities match a $\mathbf k = (1/2, 0, 1)$ ordering wavevector (antiferromagnetic modulation along the $c$-axis), with a very weak 
$\mathbf k = (1/2, 0, 0)$ also present, no modulation along the c-axis.\black

We find that the ground state magnetic order is predominantly (1/2, 0, 1) stripe antiferromagnetism, which disappears by T = 500 mK. There is a weak (1/2, 0, 0) intensity visible only at low temperature, indicating a weak canting 17.4(8) degrees away from the triangular plane, corresponding to an amplitude $0.70(4) \mu_B$ along the $z$ direction.
%
\black 
The sister compound KErSe$_2$ exhibits similar ordering, but the canting was not reported \cite{GDing2023}. \black

\section{Discussion}
\label {disc}

We have shown that a quantitative description of the field evolution of the low-lying CEF levels, including the level-crossing in the ground state, is essential for understanding the unusual properties of \cers under a magnetic field.  
To our best knowledge, such level-crossings involving the ground state and the corresponding field-induced transition have not been explicitly predicted from well-characterized CEF Hamiltonian, although it could be inferred from the phase diagram constructed from thermodynamic measurements \cite{Rotundu2004}. 
\black

This is partly attributed to the challenges in determining Stevens' coefficients in $\hcef$. Fits obtained from  zero-field spectroscopy alone often highly under constrained and suffer from substantial parameter degeneracy, especially for ions with lower site symmetry. \cite{Bordelon2020,Dun2021,Scheie_scipost2022} Here, we overcome this limitation by directly exploiting the characteristic field-dependent splitting of the Kramers doublets, thereby determining the model Hamiltonian in a robust manner. 

\begin{figure}[ht]
    \centering
    \includegraphics[width=\linewidth]{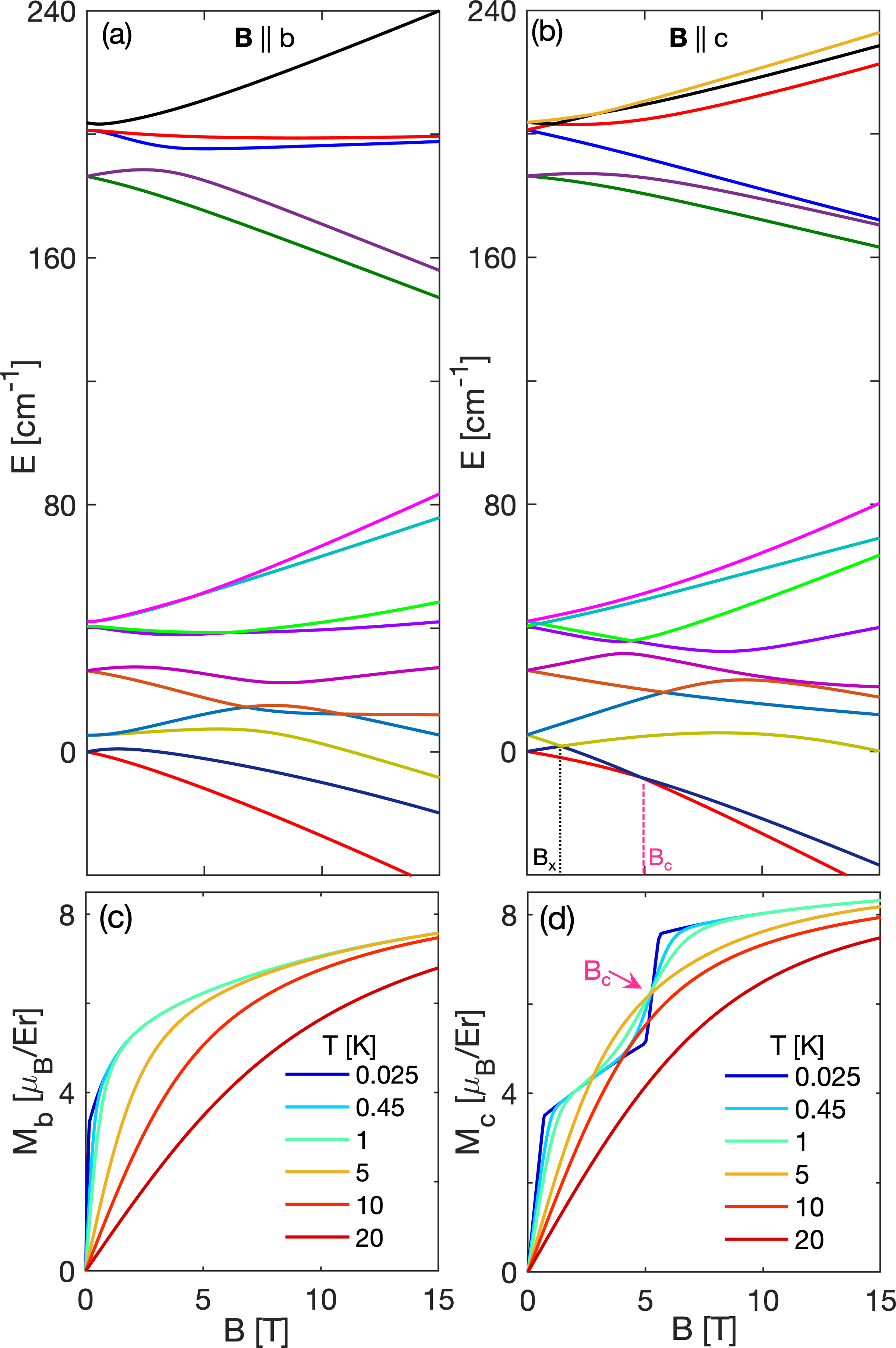}
    \caption{ Field dependent Splittings of CEF levels for the two high symmetry axes, $\Bpb$ (a) and $\Bpc$ (b). The energy levels are calculated from Eq. (\ref{eq:htot}) using the parameters shown in Table \ref{tb:bnmval}. 
    (c) and (d) display the magnetization, $M_b$ and $M_c$  as a function of applied field along the  two corresponding orientations at various temperatures as shown. }
     \label{fig:cef_full}
\end{figure}

Fig. \ref{fig:cef_full} displays all splittings of the eight doublets of \cers calculated with the parameters in Table \ref{tb:bnmval} under two high symmetry field orientations, $\Bpb$ (a) and $\Bpc$ (b).  The lowest five doublets are clustered below 40 \invcm ($\approx 5$ meV) at zero field, while the other three are clustered at around  $190 - 200$ \invcm ($\approx 25$ meV).  

The $g$-factor values are found $g_{\perp} = 6.55$ in $\Bpb$ and $g_z = 6.98$  in $\Bpc$ calculated from $g_{z(\perp)}= 2g_J|\langle 0_{\pm}|\hat J_{z(y)} |0_{\pm(\mp)}\rangle|$, where  $|0_{\pm}\rangle$ are two doubly degenerate ground states in the $B\rightarrow0$ limit.
Despite the similar $g$-factor values, the strong magnetic anisotropy of \cers is manifested by the unique field evolution of the Zeeman splitting of each doublet. 

In particular, the key distinction between Fig.~\ref{fig:cef_full}(a) and (b) lies in the ground-state evolution under applied field. For $\Bpb$, no level crossings occur among the three lowest-energy states, whereas for $\Bpc$, level crossings appear in both the ground and first excited states. These crossings are directly responsible for the unusual magnetic behavior discussed in the previous sections. Consequently, the calculated field dependences of the magnetization, $M_b$ and $M_c$ for $\Bpb$ and $\Bpc$ respectively,  differ dramatically, as shown in panels (c) and (d).

Another feature that can be revealed only by a reliable determination of the CEF parameters is the quadratic low-field field splitting of the ground state for $\Bpb$  as $B\rightarrow 0$, in contrast to the strictly linear splitting for $\Bpc$. 
This quadratic term significantly modifies the low-temperature magnetic susceptibility, especially for temperatures below the zero-field CEF gap between the two lowest doublets. 
As a result, the susceptibility deviates from the effective pseudospin-1/2 description \cite{Pocs2021}, which could be misinterpreted as other than a single-ion origin without 
an accurate understanding of the CEF spectra at finite field.

\begin{figure}[ht!]
    \includegraphics[width=\linewidth]{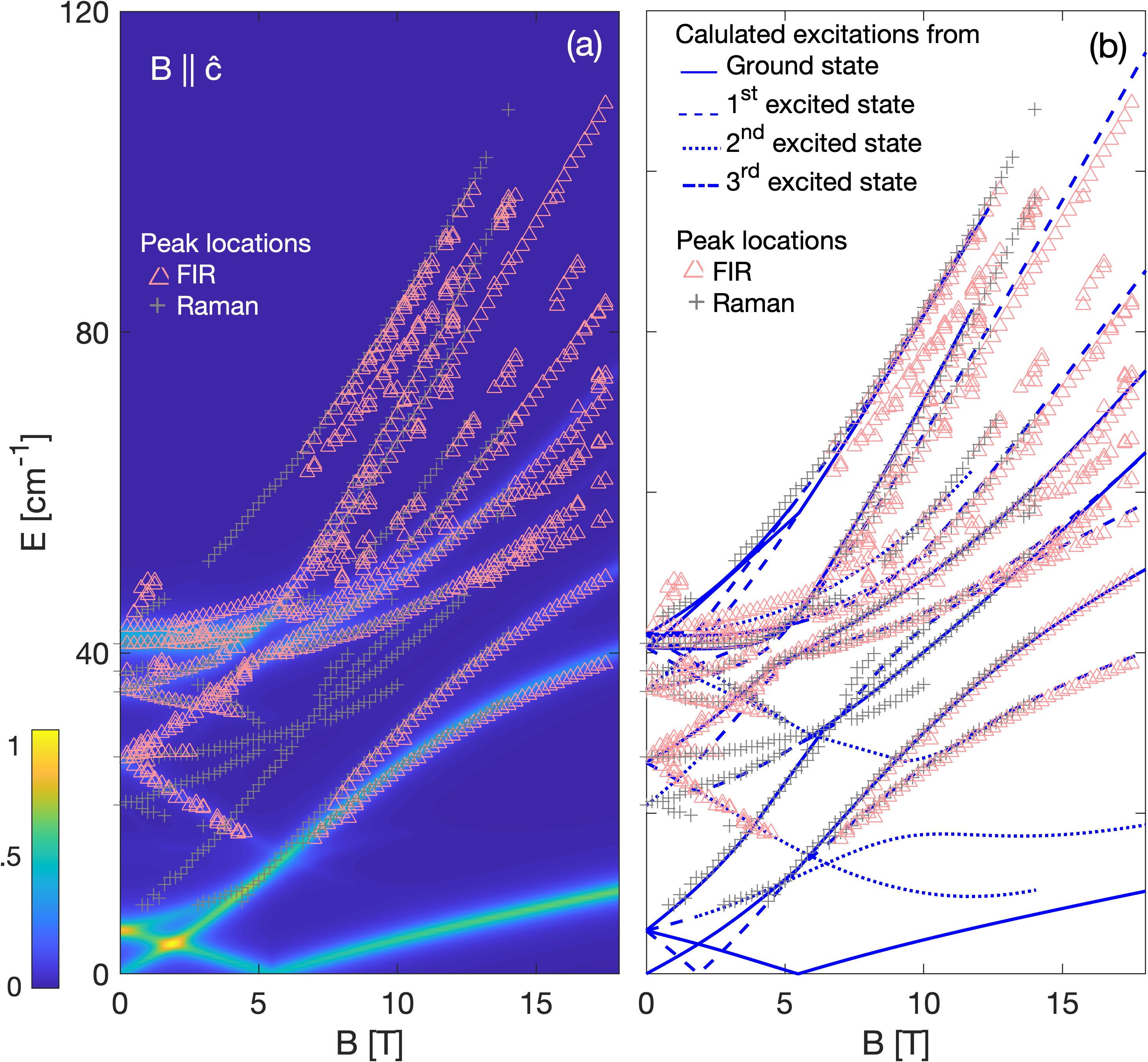}
    \caption{
    (a) Simulated field-dependent spectrum at $T=10$ K in $\Bpc$ is shown as a normalized color map.  
    Triangles (FIR)  and crosses (Raman) correspond to the peak positions from the data, Fig. \ref{fig:spec}(b) and (c), respectively.
    (b) Calculated transitions with probability $\rho> 0.001$ are shown. Solid, long-dashed, dotted  and long-short-dashed lines represent the transitions from ground, 1st, 2nd and 3rd excited states, respectively.  
    In $E>160$ \invcm, only two transitions with $\rho\ge 0.001$ emerge, which are not identifiable in the data [See Fig. \ref{fig:raman}(a)]. 
    }
    \label{fig:prob}
\end{figure}

By comparing Fig.~\ref{fig:spec}(a-c) with Fig.~\ref{fig:cef_full}(a,b), two key features of the observed CEF spectra become apparent. First, several spectral lines in Fig.~\ref{fig:spec}(a-c) cannot be explained solely by transitions originating from the ground state. Second, the three doublets with energies above 180~\invcm are not present in the measured spectra.

This is due to transitions from thermally populated low-lying excited states.  The zero-field CEF gap between the two lowest doublets is found to be 0.67 meV ($\approx$ 12 K). With an applied field, this becomes even smaller due to the level crossing in $\Bpc$, causing a finite thermal population in the excited states. 
Figure~\ref{fig:prob}(a) shows a simulated spectrum calculated using the transition probabilities from Fermi’s Golden Rule. The experimentally observed peak positions extracted from the FIR (triangles) and Raman shift (crosses) measurements in Fig.~\ref{fig:spec}(b,c) are overlaid as symbols for comparison.
We find that almost all observed peaks are well captured by transitions with a probability $\rho_c = 0.001$. 
To make a more explicit comparison, in Fig. \ref{fig:prob} (b), we plot all transitions from the three lowest states that have a probability higher than $\rho_c$ as a function of field with the same set of peak position data in panel (a), which exhibits excellent agreement. 
We also note that the discontinuities and abrupt changes in curvature of the transition lines originate from level crossings, 
where the corresponding eigenstates switch abruptly at the crossings. These changes in turn produce  discontinuous variations in the transition probabilities.

\begin{figure*}[ht]
	\includegraphics[width=\linewidth]{fig7_19may2026.png}
	\caption{(a) Raman shift ($E$) as a function of field in the larger range of Raman shift up to  $E =540$ \invcm.  
	Red lines indicate transitions with probability $\rho > 10^{-3}$, including transitions from the ground, 1st, 2nd, and 3rd excited states, the same as Fig.~\ref{fig:prob}(b).  
    The detailed features in $E<120$ \invcm~remain identical to Fig.~\ref{fig:prob}(b).\black
	Green arrows mark field-dependent peaks whereas purple arrows indicate field-independent peaks assigned to the phonon modes in Table \ref{tb:phonon}. 
	Note the intensities of any field-dependent spectra lying above 120 \invcm~are significantly smaller than those below. 
	(b) The Raman spectra intensity is shown as a function of energy  under magnetic field, varying from 0 T (bottom)  to 14 T (top) by 2 T increment.  The data are plotted with an offset for clarity.  
	Purple and green arrows  mark the same locations as panel (a).}
\label{fig:raman}
\end{figure*}

\begin{table}[b!]
\begin{center}
\caption{\small  The field-independent peaks indicated with the numbered purple arrows  Fig.~\ref{fig:raman} (a) and (b)  are listed here and compared to the phonon DFT calculation. The predicted locations' of the symmetric phonon peaks are  shown  in the parentheses in the unit of \invcm.
}
\begin{tabular*}{\linewidth}{@{\extracolsep{\fill}}ccc}
\hline\hline
&Observed [\invcm] & Assignment (calculated) \\
\hline
1&49.3(1) &$E_{2g}^1 (46.1)$\\
2&119.6(1) & unknown\\
3&125.9(1) & $E_{1g}$ (125.9)\\
4&132.7(1) &$E_{2g}^2$ (127.3)\\
5&142.8(1) &unknown\\
6&149.6(1) & $3E_{2g}^1$  \\
7&170.3(1) &$A_{1g}$ (164.8)\\
8&486.3(1) & unknown\\
9&523.8(1) & unknown\\
\hline\hline
\end{tabular*}
\label{tb:phonon}
\end{center}
\end{table}

We find that most transitions with $E>160$ cm\invcm~have calculated probabilities below $\rho_c = 0.001$, and are therefore not expected to be experimentally detectable. 
In Fig.~\ref{fig:raman}(a), we show Raman shift spectra in extended energy  up to 540 \invcm~under applied field as an intensity colormap. The red lines indicate  transitions with probability $\rho_c \geq 0.001$, identically to Fig. \ref{fig:prob}(b).   
Panel (b) displays the intensities as a function of $E$ at every 2 T increment of field. Each curve is offset by a constant for clarity.    

The green arrows of  $1',~2'$ and $3'$, mark only field-dependent features observed in $E>120$ \invcm. 
We find that none of these peak locations and intensities are matched by the transition probability as low as 10$^{-3}$, and their origins remain unclear.  Based on studies on free \erion's energy spectra \cite {Taherunnisa2019, Becker2025},  we speculate that these are related byproducts of higher order Raman processes, activated by resonance with the incoming laser energy (532 nm).  Both energies of the transitions for  $^2H_{11/2} \rightarrow ^4I_{15/2}$ (533 nm)  and   $^42S_{3/2} \rightarrow ^4I_{15/2}$ (547 nm)  lie very close to the incoming laser energy  and their intermediate transitions may cause the observed weak field dependence. 
We exclude contributions from the next $J = 13/2$ manifold, as it has been reported to lie approximately 6000 \invcm~($\approx 744$ meV) above the $J = 15/2$ manifold \cite{Becker2025}.  \black

The purple arrows $1$ to $9$  in Fig. \ref{fig:raman} indicate field-{\it{independent}} peaks, the locations of which are listed in Table \ref{tb:phonon}.  
These modes show reasonable agreement with $\Gamma$ point energies for phonon modes calculated from density functional theory (DFT). Details of the calculations are presented in \cite{supp}.  

Raman and far-infrared magneto-optical spectroscopy under intense magnetic fields offer 
a powerful and versatile approach to determining CEF physics in rare-earth materials. 
While these techniques have long been used to probe magnetic couplings and magnonic behavior in magnetic systems \cite{Ozerov2022, LXiang2023, Mou2024,Lujan2024,Mai2025}, here they are used to directly resolve the field-dependent CEF excitations in a 4f compound. This enables a precise determination of the CEF Hamiltonian in \cers, while simultaneously revealing field-induced level crossings and associated magnetic properties.

Furthermore, these methods are particularly effective for probing magnetoelastic interactions enhanced by strong spin-orbit coupling. A prominent example is the formation of vibronic bound states through hybridization between phonons and CEF levels \cite{BQLiu2018, Cermak2019}. In such cases, the resulting hybridized modes inherit the field dependence of the underlying CEF levels \cite{Ozerov2022, Mou2024}. 
Consequently,  new quasiparticles from this hybridization exhibit field-dependent spectra and dispersion, which turns out to be a natural microscopic mechanism for unusual field dependence of thermal transport phenomena \cite{Pocs2025}.

More broadly, our results demonstrate that field-dependent spectroscopy is not only a tool for determining CEF Hamiltonians, but also a powerful probe of the coupled spin-lattice physics that drives complex and unexpected magnetic behavior in rare-earth magnets.

\section{Summary}
The main result of this work is the identification of a field-induced phase transition in \ce{CsErSe2} and the associated unusual magnetic behavior emergent under field in $\Bpc$. 
This transition originates from a level crossing and subsequent ground state inversion and occurs at $\Bc\approx 5.4$ T. Such a transition is expected to produce a step-like change in magnetization and corresponding peak in magnetic susceptibility. Consistent with this expectation, our AC magnetic susceptibility measurements revealed a peak at $\Bc$, providing experimental evidence for the transition.\black
We further show that the non-monotonic evolution of the Zeeman gap below $\Bc$ is responsible for characteristic magnetic behavior, including the plateau-like field dependence of the magnetization.

This discovery was enabled by the accurate determination of our model Hamiltonian.  As in typical $4f$ electron systems, the magnetic properties of \cers are governed by single-ion physics; hence, reliably resolving the CEF scheme is essential for predicting the magnetic response.  
By directly tracking the magnetic-field dependence of the CEF level splittings using magneto-optical spectroscopy, we obtain a robust determination of the CEF Hamiltonian. 

We also find that \cers undergoes spontaneous magnetic ordering at $T_N = 110$ mK, which predominantly exhibits a stripe antiferromagnetic ordering, of which characteristics is partly attributed to dominant contributions from  $\hcef$.

The key findings of this work provide a framework for understanding unusual magnetic properties that arise from complex single-ion physics, rather than from other mechanisms such as interactions between magnetic ions. This distinction is essential for disentangling single-ion effects from collective behavior and, ultimately, for more clearly identifying the signatures of exotic spin states such as quantum spin liquids. 
\vspace{0.25in}

\begin{acknowledgments}
We thank for the fruitful discussion with Ovi Garlea,  Michael Hermele, and Martin Mourigal. 
Work at the University of Colorado Boulder was supported by 
Award No.~DE-SC0021377 of the U.S. Department of Energy (DOE), Basic Energy 
Sciences, Materials Sciences and Engineering Division (MSE). 
\noindent D.S. acknowledges support from the U. S. DOE (Grant No. DE-FG02-07ER46451) for magneto-Raman measurements.
Work at Oak Ridge National  Laboratory was supported by the U.S. DOE, MSE. 
{
	A. S. is supported by the Quantum Science Center (QSC), a National Quantum Information Science Research Center of the U.S. DOE. 
	M.O.A. acknowledges funding from the Laboratory Directed Research \& Development Program. 
The neutron scattering experiments used resources at the High Flux Isotope Reactor, a DOE Office of Science User Facility operated by the Oak Ridge National Laboratory. The beam time was allocated to HB2A on proposal number IPTS-31517.1. 
}
A part of this work was performed at the National High Magnetic Field Laboratory, which is supported by the National Science Foundation Cooperative Agreement No. DMR-1644779 and No. DMR-2128556 and the State of Florida.
\end{acknowledgments}

\vspace{0.15 in}


\begin{thebibliography}{54}%
\makeatletter
\providecommand \@ifxundefined [1]{%
 \@ifx{#1\undefined}
}%
\providecommand \@ifnum [1]{%
 \ifnum #1\expandafter \@firstoftwo
 \else \expandafter \@secondoftwo
 \fi
}%
\providecommand \@ifx [1]{%
 \ifx #1\expandafter \@firstoftwo
 \else \expandafter \@secondoftwo
 \fi
}%
\providecommand \natexlab [1]{#1}%
\providecommand \enquote  [1]{``#1''}%
\providecommand \bibnamefont  [1]{#1}%
\providecommand \bibfnamefont [1]{#1}%
\providecommand \citenamefont [1]{#1}%
\providecommand \href@noop [0]{\@secondoftwo}%
\providecommand \href [0]{\begingroup \@sanitize@url \@href}%
\providecommand \@href[1]{\@@startlink{#1}\@@href}%
\providecommand \@@href[1]{\endgroup#1\@@endlink}%
\providecommand \@sanitize@url [0]{\catcode `\\12\catcode `\$12\catcode
  `\&12\catcode `\#12\catcode `\^12\catcode `\_12\catcode `\%12\relax}%
\providecommand \@@startlink[1]{}%
\providecommand \@@endlink[0]{}%
\providecommand \url  [0]{\begingroup\@sanitize@url \@url }%
\providecommand \@url [1]{\endgroup\@href {#1}{\urlprefix }}%
\providecommand \urlprefix  [0]{URL }%
\providecommand \Eprint [0]{\href }%
\providecommand \doibase [0]{http://dx.doi.org/}%
\providecommand \selectlanguage [0]{\@gobble}%
\providecommand \bibinfo  [0]{\@secondoftwo}%
\providecommand \bibfield  [0]{\@secondoftwo}%
\providecommand \translation [1]{[#1]}%
\providecommand \BibitemOpen [0]{}%
\providecommand \bibitemStop [0]{}%
\providecommand \bibitemNoStop [0]{.\EOS\space}%
\providecommand \EOS [0]{\spacefactor3000\relax}%
\providecommand \BibitemShut  [1]{\csname bibitem#1\endcsname}%
\let\auto@bib@innerbib\@empty
\bibitem [{\citenamefont {Savary}\ and\ \citenamefont
  {Balents}(2016)}]{SavaryReview2016}%
  \BibitemOpen
  \bibfield  {author} {\bibinfo {author} {\bibfnamefont {Lucile}\ \bibnamefont
  {Savary}}\ and\ \bibinfo {author} {\bibfnamefont {Leon}\ \bibnamefont
  {Balents}},\ }\bibfield  {title} {\enquote {\bibinfo {title} {Quantum spin
  liquids: a review},}\ }\href {\doibase 10.1088/0034-4885/80/1/016502}
  {\bibfield  {journal} {\bibinfo  {journal} {Rep. Prog. Phys.}\ }\textbf
  {\bibinfo {volume} {80}},\ \bibinfo {pages} {016502} (\bibinfo {year}
  {2016})}\BibitemShut {NoStop}%
\bibitem [{\citenamefont {J.}\ \emph {et~al.}(2019)\citenamefont {J.},
  \citenamefont {Yu}, \citenamefont {Li}, \citenamefont {W.},\ and\
  \citenamefont {Li}}]{JWen2019}%
  \BibitemOpen
  \bibfield  {author} {\bibinfo {author} {\bibfnamefont {Wen}\ \bibnamefont
  {J.}}, \bibinfo {author} {\bibfnamefont {S.-L.}\ \bibnamefont {Yu}}, \bibinfo
  {author} {\bibfnamefont {S.}~\bibnamefont {Li}}, \bibinfo {author}
  {\bibfnamefont {Yu}~\bibnamefont {W.}}, \ and\ \bibinfo {author}
  {\bibfnamefont {J.-X.}\ \bibnamefont {Li}},\ }\bibfield  {title} {\enquote
  {\bibinfo {title} {Experimental identification of quantum spin liquids},}\
  }\href {\doibase 10.1038/s41535-019-0151-6} {\bibfield  {journal} {\bibinfo
  {journal} {npj Quantum Materials}\ }\textbf {\bibinfo {volume} {4}},\
  \bibinfo {pages} {12} (\bibinfo {year} {2019})}\BibitemShut {NoStop}%
\bibitem [{\citenamefont {Gardner}\ \emph {et~al.}(2010)\citenamefont
  {Gardner}, \citenamefont {Gingras},\ and\ \citenamefont
  {Greedan}}]{Gardner2010}%
  \BibitemOpen
  \bibfield  {author} {\bibinfo {author} {\bibfnamefont {Jason~S.}\
  \bibnamefont {Gardner}}, \bibinfo {author} {\bibfnamefont {Michel J.~P.}\
  \bibnamefont {Gingras}}, \ and\ \bibinfo {author} {\bibfnamefont {John~E.}\
  \bibnamefont {Greedan}},\ }\bibfield  {title} {\enquote {\bibinfo {title}
  {Magnetic pyrochlore oxides},}\ }\href {\doibase 10.1103/RevModPhys.82.53}
  {\bibfield  {journal} {\bibinfo  {journal} {Rev. Mod. Phys.}\ }\textbf
  {\bibinfo {volume} {82}},\ \bibinfo {pages} {53} (\bibinfo {year}
  {2010})}\BibitemShut {NoStop}%
\bibitem [{\citenamefont {Rau}\ and\ \citenamefont {Gingras}(2019)}]{JRau2019}%
  \BibitemOpen
  \bibfield  {author} {\bibinfo {author} {\bibfnamefont {Jeffrey~G.}\
  \bibnamefont {Rau}}\ and\ \bibinfo {author} {\bibfnamefont {Michel~J.P.}\
  \bibnamefont {Gingras}},\ }\bibfield  {title} {\enquote {\bibinfo {title}
  {{Frustrated Quantum Rare-Earth Pyrochlores}},}\ }\href {\doibase
  10.1146/annurev-conmatphys-022317-110520} {\bibfield  {journal} {\bibinfo
  {journal} {Annu. Rev. Condens. Matter Phys.}\ }\textbf {\bibinfo {volume}
  {10}},\ \bibinfo {pages} {357} (\bibinfo {year} {2019})}\BibitemShut
  {NoStop}%
\bibitem [{\citenamefont {Smith}\ \emph {et~al.}(2025)\citenamefont {Smith},
  \citenamefont {Lhotel}, \citenamefont {Petit},\ and\ \citenamefont
  {Gaulin}}]{Smith2025}%
  \BibitemOpen
  \bibfield  {author} {\bibinfo {author} {\bibfnamefont {Evan~M.}\ \bibnamefont
  {Smith}}, \bibinfo {author} {\bibfnamefont {Elsa}\ \bibnamefont {Lhotel}},
  \bibinfo {author} {\bibfnamefont {Sylvain}\ \bibnamefont {Petit}}, \ and\
  \bibinfo {author} {\bibfnamefont {Bruce~D.}\ \bibnamefont {Gaulin}},\
  }\bibfield  {title} {\enquote {\bibinfo {title} {Experimental insights into
  quantum spin ice physics in dipole-octupole pyrochlore magnets},}\ }\href
  {\doibase https://doi.org/10.1146/annurev-conmatphys-041124-015101}
  {\bibfield  {journal} {\bibinfo  {journal} {Annual Review of Condensed Matter
  Physics}\ }\textbf {\bibinfo {volume} {16}},\ \bibinfo {pages} {387--415}
  (\bibinfo {year} {2025})}\BibitemShut {NoStop}%
\bibitem [{\citenamefont {Stevens}(1952)}]{Stevens1952}%
  \BibitemOpen
  \bibfield  {author} {\bibinfo {author} {\bibfnamefont {K.~W.~H}\ \bibnamefont
  {Stevens}},\ }\bibfield  {title} {\enquote {\bibinfo {title} {Matrix elements
  and operator equivalents connected with the magnetic properties of rare earth
  ions},}\ }\href {\doibase 10.1088/0370-1298/65/3/308} {\bibfield  {journal}
  {\bibinfo  {journal} {Proc. Phys. Society. Sec. A}\ }\textbf {\bibinfo
  {volume} {65}},\ \bibinfo {pages} {209} (\bibinfo {year} {1952})}\BibitemShut
  {NoStop}%
\bibitem [{\citenamefont {Hutchings}(1964)}]{Hutchings1964}%
  \BibitemOpen
  \bibfield  {author} {\bibinfo {author} {\bibfnamefont {M.~T.}\ \bibnamefont
  {Hutchings}},\ }\bibfield  {title} {\enquote {\bibinfo {title} {Point-charge
  calculations of energy levels of magnetic ions in crystalline electric
  fields},}\ }\href {\doibase 10.1016/S0081-1947(08)60517-2} {\bibfield
  {journal} {\bibinfo  {journal} {Solid State Phys.}\ }\textbf {\bibinfo
  {volume} {16}},\ \bibinfo {pages} {227} (\bibinfo {year} {1964})}\BibitemShut
  {NoStop}%
\bibitem [{\citenamefont {Xing}\ \emph {et~al.}(2020)\citenamefont {Xing},
  \citenamefont {Sanjeewa}, \citenamefont {Kim}, \citenamefont {Stewart},
  \citenamefont {Du}, \citenamefont {Reboredo}, \citenamefont {Custelcean},\
  and\ \citenamefont {Sefat}}]{Xing2020acs}%
  \BibitemOpen
  \bibfield  {author} {\bibinfo {author} {\bibfnamefont {Jie}\ \bibnamefont
  {Xing}}, \bibinfo {author} {\bibfnamefont {Liurukara~D.}\ \bibnamefont
  {Sanjeewa}}, \bibinfo {author} {\bibfnamefont {Jungsoo}\ \bibnamefont {Kim}},
  \bibinfo {author} {\bibfnamefont {G.~R.}\ \bibnamefont {Stewart}}, \bibinfo
  {author} {\bibfnamefont {Mao-Hua}\ \bibnamefont {Du}}, \bibinfo {author}
  {\bibfnamefont {Fernando~A.}\ \bibnamefont {Reboredo}}, \bibinfo {author}
  {\bibfnamefont {Radu}\ \bibnamefont {Custelcean}}, \ and\ \bibinfo {author}
  {\bibfnamefont {Athena~S.}\ \bibnamefont {Sefat}},\ }\bibfield  {title}
  {\enquote {\bibinfo {title} {{Crystal Synthesis and Frustrated Magnetism in
  Triangular Lattice CsRESe$_2$ (RE = La--Lu): Quantum Spin Liquid Candidates
  CsCeSe$_2$ and CsYbSe$_2$}},}\ }\href {\doibase
  10.1021/acsmaterialslett.9b00464} {\bibfield  {journal} {\bibinfo  {journal}
  {ACS Mater. Lett.}\ }\textbf {\bibinfo {volume} {2}},\ \bibinfo {pages} {71}
  (\bibinfo {year} {2020})}\BibitemShut {NoStop}%
\bibitem [{\citenamefont {Bordelon}\ \emph {et~al.}(2019)\citenamefont
  {Bordelon}, \citenamefont {Kenney}, \citenamefont {Liu}, \citenamefont
  {Hogan}, \citenamefont {Posthuma}, \citenamefont {Kavand}, \citenamefont
  {Lyu}, \citenamefont {Sherwin}, \citenamefont {Butch}, \citenamefont {Brown},
  \citenamefont {Graf}, \citenamefont {Balents},\ and\ \citenamefont
  {Wilson}}]{Bordelon2019}%
  \BibitemOpen
  \bibfield  {author} {\bibinfo {author} {\bibfnamefont {Mitchell~M.}\
  \bibnamefont {Bordelon}}, \bibinfo {author} {\bibfnamefont {Eric}\
  \bibnamefont {Kenney}}, \bibinfo {author} {\bibfnamefont {Chunxiao}\
  \bibnamefont {Liu}}, \bibinfo {author} {\bibfnamefont {Tom}\ \bibnamefont
  {Hogan}}, \bibinfo {author} {\bibfnamefont {Lorenzo}\ \bibnamefont
  {Posthuma}}, \bibinfo {author} {\bibfnamefont {Marzieh}\ \bibnamefont
  {Kavand}}, \bibinfo {author} {\bibfnamefont {Yuanqi}\ \bibnamefont {Lyu}},
  \bibinfo {author} {\bibfnamefont {Mark}\ \bibnamefont {Sherwin}}, \bibinfo
  {author} {\bibfnamefont {N.~P.}\ \bibnamefont {Butch}}, \bibinfo {author}
  {\bibfnamefont {Craig}\ \bibnamefont {Brown}}, \bibinfo {author}
  {\bibfnamefont {M.~J.}\ \bibnamefont {Graf}}, \bibinfo {author}
  {\bibfnamefont {Leon}\ \bibnamefont {Balents}}, \ and\ \bibinfo {author}
  {\bibfnamefont {Stephen~D.}\ \bibnamefont {Wilson}},\ }\bibfield  {title}
  {\enquote {\bibinfo {title} {{Field-tunable quantum disordered ground state
  in the triangular-lattice antiferromagnet NaYbO$_2$}},}\ }\href {\doibase
  10.1038/s41567-019-0594-5} {\bibfield  {journal} {\bibinfo  {journal} {Nat.
  Phys.}\ }\textbf {\bibinfo {volume} {15}},\ \bibinfo {pages} {1058} (\bibinfo
  {year} {2019})}\BibitemShut {NoStop}%
\bibitem [{\citenamefont {Dai}\ \emph {et~al.}(2021)\citenamefont {Dai},
  \citenamefont {Zhang}, \citenamefont {Xie}, \citenamefont {Duan},
  \citenamefont {Gao}, \citenamefont {Zhu}, \citenamefont {Feng}, \citenamefont
  {Tao}, \citenamefont {Huang}, \citenamefont {Cao}, \citenamefont
  {Podlesnyak}, \citenamefont {Granroth}, \citenamefont {Everett},
  \citenamefont {Neuefeind}, \citenamefont {Voneshen}, \citenamefont {Wang},
  \citenamefont {Tan}, \citenamefont {Morosan}, \citenamefont {Wang},
  \citenamefont {Lin}, \citenamefont {Shu}, \citenamefont {Chen}, \citenamefont
  {Guo}, \citenamefont {Lu},\ and\ \citenamefont {Dai}}]{PDDai2021}%
  \BibitemOpen
  \bibfield  {author} {\bibinfo {author} {\bibfnamefont {Peng-Ling}\
  \bibnamefont {Dai}}, \bibinfo {author} {\bibfnamefont {Gaoning}\ \bibnamefont
  {Zhang}}, \bibinfo {author} {\bibfnamefont {Yaofeng}\ \bibnamefont {Xie}},
  \bibinfo {author} {\bibfnamefont {Chunruo}\ \bibnamefont {Duan}}, \bibinfo
  {author} {\bibfnamefont {Yonghao}\ \bibnamefont {Gao}}, \bibinfo {author}
  {\bibfnamefont {Zihao}\ \bibnamefont {Zhu}}, \bibinfo {author} {\bibfnamefont
  {Erxi}\ \bibnamefont {Feng}}, \bibinfo {author} {\bibfnamefont {Zhen}\
  \bibnamefont {Tao}}, \bibinfo {author} {\bibfnamefont {Chien-Lung}\
  \bibnamefont {Huang}}, \bibinfo {author} {\bibfnamefont {Huibo}\ \bibnamefont
  {Cao}}, \bibinfo {author} {\bibfnamefont {Andrey}\ \bibnamefont
  {Podlesnyak}}, \bibinfo {author} {\bibfnamefont {Garrett~E.}\ \bibnamefont
  {Granroth}}, \bibinfo {author} {\bibfnamefont {Michelle~S.}\ \bibnamefont
  {Everett}}, \bibinfo {author} {\bibfnamefont {Joerg~C.}\ \bibnamefont
  {Neuefeind}}, \bibinfo {author} {\bibfnamefont {David}\ \bibnamefont
  {Voneshen}}, \bibinfo {author} {\bibfnamefont {Shun}\ \bibnamefont {Wang}},
  \bibinfo {author} {\bibfnamefont {Guotai}\ \bibnamefont {Tan}}, \bibinfo
  {author} {\bibfnamefont {Emilia}\ \bibnamefont {Morosan}}, \bibinfo {author}
  {\bibfnamefont {Xia}\ \bibnamefont {Wang}}, \bibinfo {author} {\bibfnamefont
  {Hai-Qing}\ \bibnamefont {Lin}}, \bibinfo {author} {\bibfnamefont {Lei}\
  \bibnamefont {Shu}}, \bibinfo {author} {\bibfnamefont {Gang}\ \bibnamefont
  {Chen}}, \bibinfo {author} {\bibfnamefont {Yanfeng}\ \bibnamefont {Guo}},
  \bibinfo {author} {\bibfnamefont {Xingye}\ \bibnamefont {Lu}}, \ and\
  \bibinfo {author} {\bibfnamefont {Pengcheng}\ \bibnamefont {Dai}},\
  }\bibfield  {title} {\enquote {\bibinfo {title} {{Spinon Fermi Surface Spin
  Liquid in a Triangular Lattice Antiferromagnet ${\mathrm{NaYbSe}}_{2}$}},}\
  }\href {\doibase 10.1103/PhysRevX.11.021044} {\bibfield  {journal} {\bibinfo
  {journal} {Phys. Rev. X}\ }\textbf {\bibinfo {volume} {11}},\ \bibinfo
  {pages} {021044} (\bibinfo {year} {2021})}\BibitemShut {NoStop}%
\bibitem [{\citenamefont {Zhang}\ \emph {et~al.}(2021)\citenamefont {Zhang},
  \citenamefont {Ma}, \citenamefont {Li}, \citenamefont {Wang}, \citenamefont
  {Adroja}, \citenamefont {Perring}, \citenamefont {Liu}, \citenamefont {Jin},
  \citenamefont {Ji}, \citenamefont {Wang}, \citenamefont {Kamiya},
  \citenamefont {Wang}, \citenamefont {Ma},\ and\ \citenamefont
  {Zhang}}]{Zhang2021}%
  \BibitemOpen
  \bibfield  {author} {\bibinfo {author} {\bibfnamefont {Zheng}\ \bibnamefont
  {Zhang}}, \bibinfo {author} {\bibfnamefont {Xiaoli}\ \bibnamefont {Ma}},
  \bibinfo {author} {\bibfnamefont {Jianshu}\ \bibnamefont {Li}}, \bibinfo
  {author} {\bibfnamefont {Guohua}\ \bibnamefont {Wang}}, \bibinfo {author}
  {\bibfnamefont {D.~T.}\ \bibnamefont {Adroja}}, \bibinfo {author}
  {\bibfnamefont {T.~P.}\ \bibnamefont {Perring}}, \bibinfo {author}
  {\bibfnamefont {Weiwei}\ \bibnamefont {Liu}}, \bibinfo {author}
  {\bibfnamefont {Feng}\ \bibnamefont {Jin}}, \bibinfo {author} {\bibfnamefont
  {Jianting}\ \bibnamefont {Ji}}, \bibinfo {author} {\bibfnamefont {Yimeng}\
  \bibnamefont {Wang}}, \bibinfo {author} {\bibfnamefont {Yoshitomo}\
  \bibnamefont {Kamiya}}, \bibinfo {author} {\bibfnamefont {Xiaoqun}\
  \bibnamefont {Wang}}, \bibinfo {author} {\bibfnamefont {Jie}\ \bibnamefont
  {Ma}}, \ and\ \bibinfo {author} {\bibfnamefont {Qingming}\ \bibnamefont
  {Zhang}},\ }\bibfield  {title} {\enquote {\bibinfo {title} {{Crystalline
  electric field excitations in the quantum spin liquid candidate
  NaYbSe$_2$}},}\ }\href {\doibase 10.1103/PhysRevB.103.035144} {\bibfield
  {journal} {\bibinfo  {journal} {Phys. Rev. B}\ }\textbf {\bibinfo {volume}
  {103}},\ \bibinfo {pages} {035144} (\bibinfo {year} {2021})}\BibitemShut
  {NoStop}%
\bibitem [{\citenamefont {Pocs}\ \emph {et~al.}(2021)\citenamefont {Pocs},
  \citenamefont {Siegfried}, \citenamefont {Xing}, \citenamefont {Sefat},
  \citenamefont {Hermele}, \citenamefont {Normand},\ and\ \citenamefont
  {Lee}}]{Pocs2021}%
  \BibitemOpen
  \bibfield  {author} {\bibinfo {author} {\bibfnamefont {Christopher~A.}\
  \bibnamefont {Pocs}}, \bibinfo {author} {\bibfnamefont {Peter~E.}\
  \bibnamefont {Siegfried}}, \bibinfo {author} {\bibfnamefont {Jie}\
  \bibnamefont {Xing}}, \bibinfo {author} {\bibfnamefont {Athena~S.}\
  \bibnamefont {Sefat}}, \bibinfo {author} {\bibfnamefont {Michael}\
  \bibnamefont {Hermele}}, \bibinfo {author} {\bibfnamefont {B.}~\bibnamefont
  {Normand}}, \ and\ \bibinfo {author} {\bibfnamefont {Minhyea}\ \bibnamefont
  {Lee}},\ }\bibfield  {title} {\enquote {\bibinfo {title} {Systematic
  extraction of crystal electric-field effects and quantum magnetic model
  parameters in triangular rare-earth magnets},}\ }\href {\doibase
  10.1103/PhysRevResearch.3.043202} {\bibfield  {journal} {\bibinfo  {journal}
  {Phys. Rev. Res.}\ }\textbf {\bibinfo {volume} {3}},\ \bibinfo {pages}
  {043202} (\bibinfo {year} {2021})}\BibitemShut {NoStop}%
\bibitem [{\citenamefont {Scheie}\ \emph {et~al.}(2024)\citenamefont {Scheie},
  \citenamefont {Ghioldi}, \citenamefont {Xing}, \citenamefont {Paddison},
  \citenamefont {Sherman}, \citenamefont {Dupont}, \citenamefont {Sanjeewa},
  \citenamefont {Lee}, \citenamefont {Woods}, \citenamefont {Abernathy},
  \citenamefont {Pajerowski}, \citenamefont {Williams}, \citenamefont {Zhang},
  \citenamefont {Manuel}, \citenamefont {Trumper}, \citenamefont {Pemmaraju},
  \citenamefont {Sefat}, \citenamefont {Parker}, \citenamefont {Devereaux},
  \citenamefont {Movshovich}, \citenamefont {Moore}, \citenamefont {Batista},\
  and\ \citenamefont {Tennant}}]{Scheie2024}%
  \BibitemOpen
  \bibfield  {author} {\bibinfo {author} {\bibfnamefont {A.~O.}\ \bibnamefont
  {Scheie}}, \bibinfo {author} {\bibfnamefont {E.~A.}\ \bibnamefont {Ghioldi}},
  \bibinfo {author} {\bibfnamefont {J.}~\bibnamefont {Xing}}, \bibinfo {author}
  {\bibfnamefont {J.~A.~M.}\ \bibnamefont {Paddison}}, \bibinfo {author}
  {\bibfnamefont {N.~E.}\ \bibnamefont {Sherman}}, \bibinfo {author}
  {\bibfnamefont {M.}~\bibnamefont {Dupont}}, \bibinfo {author} {\bibfnamefont
  {L.~D.}\ \bibnamefont {Sanjeewa}}, \bibinfo {author} {\bibfnamefont
  {Sangyun}\ \bibnamefont {Lee}}, \bibinfo {author} {\bibfnamefont {A.~J.}\
  \bibnamefont {Woods}}, \bibinfo {author} {\bibfnamefont {D.}~\bibnamefont
  {Abernathy}}, \bibinfo {author} {\bibfnamefont {D.~M.}\ \bibnamefont
  {Pajerowski}}, \bibinfo {author} {\bibfnamefont {T.~J.}\ \bibnamefont
  {Williams}}, \bibinfo {author} {\bibfnamefont {Shang-Shun}\ \bibnamefont
  {Zhang}}, \bibinfo {author} {\bibfnamefont {L.~O.}\ \bibnamefont {Manuel}},
  \bibinfo {author} {\bibfnamefont {A.~E.}\ \bibnamefont {Trumper}}, \bibinfo
  {author} {\bibfnamefont {C.~D.}\ \bibnamefont {Pemmaraju}}, \bibinfo {author}
  {\bibfnamefont {A.~S.}\ \bibnamefont {Sefat}}, \bibinfo {author}
  {\bibfnamefont {D.~S.}\ \bibnamefont {Parker}}, \bibinfo {author}
  {\bibfnamefont {T.~P.}\ \bibnamefont {Devereaux}}, \bibinfo {author}
  {\bibfnamefont {R.}~\bibnamefont {Movshovich}}, \bibinfo {author}
  {\bibfnamefont {J.~E.}\ \bibnamefont {Moore}}, \bibinfo {author}
  {\bibfnamefont {C.~D.}\ \bibnamefont {Batista}}, \ and\ \bibinfo {author}
  {\bibfnamefont {D.~A.}\ \bibnamefont {Tennant}},\ }\bibfield  {title}
  {\enquote {\bibinfo {title} {Proximate spin liquid and fractionalization in
  the triangular antiferromagnet kybse2},}\ }\href {\doibase
  10.1038/s41567-023-02259-1} {\bibfield  {journal} {\bibinfo  {journal} {Nat.
  Phys}\ }\textbf {\bibinfo {volume} {20}},\ \bibinfo {pages} {74--81}
  (\bibinfo {year} {2024})}\BibitemShut {NoStop}%
\bibitem [{\citenamefont {Scheie}\ \emph {et~al.}(2020)\citenamefont {Scheie},
  \citenamefont {Garlea}, \citenamefont {Sanjeewa}, \citenamefont {Xing},\ and\
  \citenamefont {Sefat}}]{Scheie2020Er}%
  \BibitemOpen
  \bibfield  {author} {\bibinfo {author} {\bibfnamefont {A.}~\bibnamefont
  {Scheie}}, \bibinfo {author} {\bibfnamefont {V.~O.}\ \bibnamefont {Garlea}},
  \bibinfo {author} {\bibfnamefont {L.~D.}\ \bibnamefont {Sanjeewa}}, \bibinfo
  {author} {\bibfnamefont {J.}~\bibnamefont {Xing}}, \ and\ \bibinfo {author}
  {\bibfnamefont {A.~S.}\ \bibnamefont {Sefat}},\ }\bibfield  {title} {\enquote
  {\bibinfo {title} {Crystal-field hamiltonian and anisotropy in
  ${\mathrm{kerse}}_{2}$ and ${\mathrm{cserse}}_{2}$},}\ }\href {\doibase
  10.1103/PhysRevB.101.144432} {\bibfield  {journal} {\bibinfo  {journal}
  {Phys. Rev. B}\ }\textbf {\bibinfo {volume} {101}},\ \bibinfo {pages}
  {144432} (\bibinfo {year} {2020})}\BibitemShut {NoStop}%
\bibitem [{\citenamefont {Xing}\ \emph {et~al.}(2019)\citenamefont {Xing},
  \citenamefont {Sanjeewa}, \citenamefont {Kim}, \citenamefont {Meier},
  \citenamefont {May}, \citenamefont {Zheng}, \citenamefont {Custelcean},
  \citenamefont {Stewart},\ and\ \citenamefont {Sefat}}]{Xing2019prm}%
  \BibitemOpen
  \bibfield  {author} {\bibinfo {author} {\bibfnamefont {Jie}\ \bibnamefont
  {Xing}}, \bibinfo {author} {\bibfnamefont {Liurukara~D.}\ \bibnamefont
  {Sanjeewa}}, \bibinfo {author} {\bibfnamefont {Jungsoo}\ \bibnamefont {Kim}},
  \bibinfo {author} {\bibfnamefont {William~R.}\ \bibnamefont {Meier}},
  \bibinfo {author} {\bibfnamefont {Andrew~F.}\ \bibnamefont {May}}, \bibinfo
  {author} {\bibfnamefont {Qiang}\ \bibnamefont {Zheng}}, \bibinfo {author}
  {\bibfnamefont {Radu}\ \bibnamefont {Custelcean}}, \bibinfo {author}
  {\bibfnamefont {G.~R.}\ \bibnamefont {Stewart}}, \ and\ \bibinfo {author}
  {\bibfnamefont {Athena~S.}\ \bibnamefont {Sefat}},\ }\bibfield  {title}
  {\enquote {\bibinfo {title} {Synthesis, magnetization, and heat capacity of
  triangular lattice materials ${\mathrm{naerse}}_{2}$ and
  ${\mathrm{kerse}}_{2}$},}\ }\href {\doibase
  10.1103/PhysRevMaterials.3.114413} {\bibfield  {journal} {\bibinfo  {journal}
  {Phys. Rev. Mater.}\ }\textbf {\bibinfo {volume} {3}},\ \bibinfo {pages}
  {114413} (\bibinfo {year} {2019})}\BibitemShut {NoStop}%
\bibitem [{\citenamefont {Ding}\ \emph {et~al.}(2023)\citenamefont {Ding},
  \citenamefont {Wo}, \citenamefont {Luo}, \citenamefont {Gu}, \citenamefont
  {Gu}, \citenamefont {Bewley}, \citenamefont {Chen},\ and\ \citenamefont
  {Zhao}}]{GDing2023}%
  \BibitemOpen
  \bibfield  {author} {\bibinfo {author} {\bibfnamefont {Gaofeng}\ \bibnamefont
  {Ding}}, \bibinfo {author} {\bibfnamefont {Hongliang}\ \bibnamefont {Wo}},
  \bibinfo {author} {\bibfnamefont {Rui~Leonard}\ \bibnamefont {Luo}}, \bibinfo
  {author} {\bibfnamefont {Yimeng}\ \bibnamefont {Gu}}, \bibinfo {author}
  {\bibfnamefont {Yiqing}\ \bibnamefont {Gu}}, \bibinfo {author} {\bibfnamefont
  {Robert}\ \bibnamefont {Bewley}}, \bibinfo {author} {\bibfnamefont {Gang}\
  \bibnamefont {Chen}}, \ and\ \bibinfo {author} {\bibfnamefont {Jun}\
  \bibnamefont {Zhao}},\ }\bibfield  {title} {\enquote {\bibinfo {title}
  {Stripe order and spin dynamics in the triangular-lattice antiferromagnet
  ${\mathrm{kerse}}_{2}$: A single-crystal study with a theoretical
  description},}\ }\href {\doibase 10.1103/PhysRevB.107.L100411} {\bibfield
  {journal} {\bibinfo  {journal} {Phys. Rev. B}\ }\textbf {\bibinfo {volume}
  {107}},\ \bibinfo {pages} {L100411} (\bibinfo {year} {2023})}\BibitemShut
  {NoStop}%
\bibitem [{\citenamefont {Garlea}\ \emph {et~al.}(2010)\citenamefont {Garlea},
  \citenamefont {Chakoumakos}, \citenamefont {Moore}, \citenamefont {Taylor},
  \citenamefont {Chae}, \citenamefont {Maples}, \citenamefont {Riedel},
  \citenamefont {Lynn},\ and\ \citenamefont {Selby}}]{garlea2010high}%
  \BibitemOpen
  \bibfield  {author} {\bibinfo {author} {\bibfnamefont {Vasile~O}\
  \bibnamefont {Garlea}}, \bibinfo {author} {\bibfnamefont {Bryan~C}\
  \bibnamefont {Chakoumakos}}, \bibinfo {author} {\bibfnamefont {Scott~A}\
  \bibnamefont {Moore}}, \bibinfo {author} {\bibfnamefont {Gerald~Brent}\
  \bibnamefont {Taylor}}, \bibinfo {author} {\bibfnamefont {Timothy}\
  \bibnamefont {Chae}}, \bibinfo {author} {\bibfnamefont {Ron~G}\ \bibnamefont
  {Maples}}, \bibinfo {author} {\bibfnamefont {Richard~A}\ \bibnamefont
  {Riedel}}, \bibinfo {author} {\bibfnamefont {Gary~W}\ \bibnamefont {Lynn}}, \
  and\ \bibinfo {author} {\bibfnamefont {Douglas~L}\ \bibnamefont {Selby}},\
  }\bibfield  {title} {\enquote {\bibinfo {title} {The high-resolution powder
  diffractometer at the high flux isotope reactor},}\ }\href@noop {} {\bibfield
   {journal} {\bibinfo  {journal} {Applied Physics A}\ }\textbf {\bibinfo
  {volume} {99}},\ \bibinfo {pages} {531--535} (\bibinfo {year}
  {2010})}\BibitemShut {NoStop}%
\bibitem [{\citenamefont
  {Rodr{\'\i}guez-Carvajal}(1993)}]{rodriguez1993recent}%
  \BibitemOpen
  \bibfield  {author} {\bibinfo {author} {\bibfnamefont {Juan}\ \bibnamefont
  {Rodr{\'\i}guez-Carvajal}},\ }\bibfield  {title} {\enquote {\bibinfo {title}
  {Recent advances in magnetic structure determination by neutron powder
  diffraction},}\ }\href@noop {} {\bibfield  {journal} {\bibinfo  {journal}
  {Physica B: Condensed Matter}\ }\textbf {\bibinfo {volume} {192}},\ \bibinfo
  {pages} {55--69} (\bibinfo {year} {1993})}\BibitemShut {NoStop}%
\bibitem [{\citenamefont {Kresse}\ and\ \citenamefont
  {Hafner}(1993)}]{kresse1993ab}%
  \BibitemOpen
  \bibfield  {author} {\bibinfo {author} {\bibfnamefont {Georg}\ \bibnamefont
  {Kresse}}\ and\ \bibinfo {author} {\bibfnamefont {J{\"u}rgen}\ \bibnamefont
  {Hafner}},\ }\bibfield  {title} {\enquote {\bibinfo {title} {Ab initio
  molecular dynamics for liquid metals},}\ }\href@noop {} {\bibfield  {journal}
  {\bibinfo  {journal} {{Phys. Rev. B}}\ }\textbf {\bibinfo {volume} {47}},\
  \bibinfo {pages} {558} (\bibinfo {year} {1993})}\BibitemShut {NoStop}%
\bibitem [{\citenamefont {Kresse}\ and\ \citenamefont
  {Furthm{\"u}ller}(1996{\natexlab{a}})}]{kresse1996efficient}%
  \BibitemOpen
  \bibfield  {author} {\bibinfo {author} {\bibfnamefont {Georg}\ \bibnamefont
  {Kresse}}\ and\ \bibinfo {author} {\bibfnamefont {J{\"u}rgen}\ \bibnamefont
  {Furthm{\"u}ller}},\ }\bibfield  {title} {\enquote {\bibinfo {title}
  {Efficient iterative schemes for \emph{ab initio} total-energy calculations
  using a plane-wave basis set},}\ }\href@noop {} {\bibfield  {journal}
  {\bibinfo  {journal} {{Phys. Rev. B}}\ }\textbf {\bibinfo {volume} {54}},\
  \bibinfo {pages} {11169} (\bibinfo {year} {1996}{\natexlab{a}})}\BibitemShut
  {NoStop}%
\bibitem [{\citenamefont {Kresse}\ and\ \citenamefont
  {Furthm{\"u}ller}(1996{\natexlab{b}})}]{kresse1996efficiency}%
  \BibitemOpen
  \bibfield  {author} {\bibinfo {author} {\bibfnamefont {Georg}\ \bibnamefont
  {Kresse}}\ and\ \bibinfo {author} {\bibfnamefont {J{\"u}rgen}\ \bibnamefont
  {Furthm{\"u}ller}},\ }\bibfield  {title} {\enquote {\bibinfo {title}
  {Efficiency of ab-initio total energy calculations for metals and
  semiconductors using a plane-wave basis set},}\ }\href@noop {} {\bibfield
  {journal} {\bibinfo  {journal} {{Comput. Mater. Sci.}}\ }\textbf {\bibinfo
  {volume} {6}},\ \bibinfo {pages} {15--50} (\bibinfo {year}
  {1996}{\natexlab{b}})}\BibitemShut {NoStop}%
\bibitem [{\citenamefont {Kresse}\ and\ \citenamefont
  {Hafner}(1994)}]{kresse1994norm}%
  \BibitemOpen
  \bibfield  {author} {\bibinfo {author} {\bibfnamefont {G}~\bibnamefont
  {Kresse}}\ and\ \bibinfo {author} {\bibfnamefont {J}~\bibnamefont {Hafner}},\
  }\bibfield  {title} {\enquote {\bibinfo {title} {{Norm-Conserving and
  Ultrasoft Pseudopotentials for First-Row and Transition Elements}},}\
  }\href@noop {} {\bibfield  {journal} {\bibinfo  {journal} {{J. Phys.:
  Condens.Matter}}\ }\textbf {\bibinfo {volume} {6}},\ \bibinfo {pages} {8245}
  (\bibinfo {year} {1994})}\BibitemShut {NoStop}%
\bibitem [{\citenamefont {Bl\"ochl}(1994)}]{blochl1994projector}%
  \BibitemOpen
  \bibfield  {author} {\bibinfo {author} {\bibfnamefont {P.~E.}\ \bibnamefont
  {Bl\"ochl}},\ }\bibfield  {title} {\enquote {\bibinfo {title} {{Projector
  Augmented-Wave Method}},}\ }\href {\doibase 10.1103/PhysRevB.50.17953}
  {\bibfield  {journal} {\bibinfo  {journal} {{Phys. Rev. B}}\ }\textbf
  {\bibinfo {volume} {50}},\ \bibinfo {pages} {17953--17979} (\bibinfo {year}
  {1994})}\BibitemShut {NoStop}%
\bibitem [{\citenamefont {Kresse}\ and\ \citenamefont
  {Joubert}(1999)}]{kresse1999ultrasoft}%
  \BibitemOpen
  \bibfield  {author} {\bibinfo {author} {\bibfnamefont {Georg}\ \bibnamefont
  {Kresse}}\ and\ \bibinfo {author} {\bibfnamefont {Daniel}\ \bibnamefont
  {Joubert}},\ }\bibfield  {title} {\enquote {\bibinfo {title} {{From Ultrasoft
  Pseudopotentials to the Projector Augmented-Wave Method}},}\ }\href@noop {}
  {\bibfield  {journal} {\bibinfo  {journal} {{Phys. Rev. B}}\ }\textbf
  {\bibinfo {volume} {59}},\ \bibinfo {pages} {1758} (\bibinfo {year}
  {1999})}\BibitemShut {NoStop}%
\bibitem [{\citenamefont {Perdew}\ \emph {et~al.}(1996)\citenamefont {Perdew},
  \citenamefont {Burke},\ and\ \citenamefont
  {Ernzerhof}}]{perdew1996generalized}%
  \BibitemOpen
  \bibfield  {author} {\bibinfo {author} {\bibfnamefont {John~P}\ \bibnamefont
  {Perdew}}, \bibinfo {author} {\bibfnamefont {Kieron}\ \bibnamefont {Burke}},
  \ and\ \bibinfo {author} {\bibfnamefont {Matthias}\ \bibnamefont
  {Ernzerhof}},\ }\bibfield  {title} {\enquote {\bibinfo {title} {{Generalized
  Gradient Approximation Made Simple}},}\ }\href@noop {} {\bibfield  {journal}
  {\bibinfo  {journal} {{Phys. Rev. Lett.}}\ }\textbf {\bibinfo {volume}
  {77}},\ \bibinfo {pages} {3865} (\bibinfo {year} {1996})}\BibitemShut
  {NoStop}%
\bibitem [{\citenamefont {Monkhorst}\ and\ \citenamefont
  {Pack}(1976)}]{PhysRevB.13.5188}%
  \BibitemOpen
  \bibfield  {author} {\bibinfo {author} {\bibfnamefont {Hendrik~J.}\
  \bibnamefont {Monkhorst}}\ and\ \bibinfo {author} {\bibfnamefont {James~D.}\
  \bibnamefont {Pack}},\ }\bibfield  {title} {\enquote {\bibinfo {title}
  {Special points for brillouin-zone integrations},}\ }\href {\doibase
  10.1103/PhysRevB.13.5188} {\bibfield  {journal} {\bibinfo  {journal} {Phys.
  Rev. B}\ }\textbf {\bibinfo {volume} {13}},\ \bibinfo {pages} {5188--5192}
  (\bibinfo {year} {1976})}\BibitemShut {NoStop}%
\bibitem [{\citenamefont {Togo}\ \emph {et~al.}(2023)\citenamefont {Togo},
  \citenamefont {Chaput}, \citenamefont {Tadano},\ and\ \citenamefont
  {Tanaka}}]{phonopy-phono3py-JPCM}%
  \BibitemOpen
  \bibfield  {author} {\bibinfo {author} {\bibfnamefont {Atsushi}\ \bibnamefont
  {Togo}}, \bibinfo {author} {\bibfnamefont {Laurent}\ \bibnamefont {Chaput}},
  \bibinfo {author} {\bibfnamefont {Terumasa}\ \bibnamefont {Tadano}}, \ and\
  \bibinfo {author} {\bibfnamefont {Isao}\ \bibnamefont {Tanaka}},\ }\bibfield
  {title} {\enquote {\bibinfo {title} {Implementation strategies in phonopy and
  phono3py},}\ }\href {\doibase 10.1088/1361-648X/acd831} {\bibfield  {journal}
  {\bibinfo  {journal} {J. Phys. Condens. Matter}\ }\textbf {\bibinfo {volume}
  {35}},\ \bibinfo {pages} {353001} (\bibinfo {year} {2023})}\BibitemShut
  {NoStop}%
\bibitem [{\citenamefont {Togo}(2023)}]{phonopy-phono3py-JPSJ}%
  \BibitemOpen
  \bibfield  {author} {\bibinfo {author} {\bibfnamefont {Atsushi}\ \bibnamefont
  {Togo}},\ }\bibfield  {title} {\enquote {\bibinfo {title} {First-principles
  phonon calculations with phonopy and phono3py},}\ }\href {\doibase
  10.7566/JPSJ.92.012001} {\bibfield  {journal} {\bibinfo  {journal} {J. Phys.
  Soc. Jpn.}\ }\textbf {\bibinfo {volume} {92}},\ \bibinfo {pages} {012001}
  (\bibinfo {year} {2023})}\BibitemShut {NoStop}%
\bibitem [{\citenamefont {Yamamoto}\ \emph {et~al.}(2014)\citenamefont
  {Yamamoto}, \citenamefont {Marmorini},\ and\ \citenamefont
  {Danshita}}]{Yamamoto2014}%
  \BibitemOpen
  \bibfield  {author} {\bibinfo {author} {\bibfnamefont {Daisuke}\ \bibnamefont
  {Yamamoto}}, \bibinfo {author} {\bibfnamefont {Giacomo}\ \bibnamefont
  {Marmorini}}, \ and\ \bibinfo {author} {\bibfnamefont {Ippei}\ \bibnamefont
  {Danshita}},\ }\bibfield  {title} {\enquote {\bibinfo {title} {{Quantum Phase
  Diagram of the Triangular-Lattice $XXZ$ Model in a Magnetic Field}},}\ }\href
  {\doibase 10.1103/PhysRevLett.112.127203} {\bibfield  {journal} {\bibinfo
  {journal} {Phys. Rev. Lett.}\ }\textbf {\bibinfo {volume} {112}},\ \bibinfo
  {pages} {127203} (\bibinfo {year} {2014})}\BibitemShut {NoStop}%
\bibitem [{\citenamefont {Maksimov}\ \emph {et~al.}(2019)\citenamefont
  {Maksimov}, \citenamefont {Zhu}, \citenamefont {White},\ and\ \citenamefont
  {Chernyshev}}]{Maksimov2019}%
  \BibitemOpen
  \bibfield  {author} {\bibinfo {author} {\bibfnamefont {P.~A.}\ \bibnamefont
  {Maksimov}}, \bibinfo {author} {\bibfnamefont {Zhenyue}\ \bibnamefont {Zhu}},
  \bibinfo {author} {\bibfnamefont {Steven~R.}\ \bibnamefont {White}}, \ and\
  \bibinfo {author} {\bibfnamefont {A.~L.}\ \bibnamefont {Chernyshev}},\
  }\bibfield  {title} {\enquote {\bibinfo {title} {{Anisotropic-Exchange
  Magnets on a Triangular Lattice: Spin Waves, Accidental Degeneracies, and
  Dual Spin Liquids}},}\ }\href {\doibase 10.1103/PhysRevX.9.021017} {\bibfield
   {journal} {\bibinfo  {journal} {Phys. Rev. X}\ }\textbf {\bibinfo {volume}
  {9}},\ \bibinfo {pages} {021017} (\bibinfo {year} {2019})}\BibitemShut
  {NoStop}%
\bibitem [{not()}]{note1}%
  \BibitemOpen
  \href@noop {} {}\bibinfo {note} {We note that the quantization axis defining
  the $\hat{z}$ direction of the chosen basis of spin operators in
  $\mathcal{H}_{\rm CEF}$ and the spin interaction term of $\mathcal{H}_{\rm
  XXZ}$ is identically the crystallographic $c$-axis, whence the component
  $B_z$ refers to a field ($\mathbf B \parallel c$) applied along the
  crystalline $c$-axis in experiment, and $B_y$ to the field applied along the
  $b$-axis within the $ab$-plane.}\BibitemShut {Stop}%
\bibitem [{\citenamefont {Virtanen}\ \emph {et~al.}(2020)\citenamefont
  {Virtanen}, \citenamefont {Gommers}, \citenamefont {Oliphant}, \citenamefont
  {Haberland}, \citenamefont {Reddy}, \citenamefont {Cournapeau}, \citenamefont
  {Burovski}, \citenamefont {Peterson}, \citenamefont {Weckesser},
  \citenamefont {Bright}, \citenamefont {{van der Walt}}, \citenamefont
  {Brett}, \citenamefont {Wilson}, \citenamefont {Millman}, \citenamefont
  {Mayorov}, \citenamefont {Nelson}, \citenamefont {Jones}, \citenamefont
  {Kern}, \citenamefont {Larson}, \citenamefont {Carey}, \citenamefont {Polat},
  \citenamefont {Feng}, \citenamefont {Moore}, \citenamefont {{VanderPlas}},
  \citenamefont {Laxalde}, \citenamefont {Perktold}, \citenamefont {Cimrman},
  \citenamefont {Henriksen}, \citenamefont {Quintero}, \citenamefont {Harris},
  \citenamefont {Archibald}, \citenamefont {Ribeiro}, \citenamefont
  {Pedregosa}, \citenamefont {{van Mulbregt}},\ and\ \citenamefont {{SciPy 1.0
  Contributors}}}]{2020SciPy-NMeth}%
  \BibitemOpen
  \bibfield  {author} {\bibinfo {author} {\bibfnamefont {Pauli}\ \bibnamefont
  {Virtanen}}, \bibinfo {author} {\bibfnamefont {Ralf}\ \bibnamefont
  {Gommers}}, \bibinfo {author} {\bibfnamefont {Travis~E.}\ \bibnamefont
  {Oliphant}}, \bibinfo {author} {\bibfnamefont {Matt}\ \bibnamefont
  {Haberland}}, \bibinfo {author} {\bibfnamefont {Tyler}\ \bibnamefont
  {Reddy}}, \bibinfo {author} {\bibfnamefont {David}\ \bibnamefont
  {Cournapeau}}, \bibinfo {author} {\bibfnamefont {Evgeni}\ \bibnamefont
  {Burovski}}, \bibinfo {author} {\bibfnamefont {Pearu}\ \bibnamefont
  {Peterson}}, \bibinfo {author} {\bibfnamefont {Warren}\ \bibnamefont
  {Weckesser}}, \bibinfo {author} {\bibfnamefont {Jonathan}\ \bibnamefont
  {Bright}}, \bibinfo {author} {\bibfnamefont {St{\'e}fan~J.}\ \bibnamefont
  {{van der Walt}}}, \bibinfo {author} {\bibfnamefont {Matthew}\ \bibnamefont
  {Brett}}, \bibinfo {author} {\bibfnamefont {Joshua}\ \bibnamefont {Wilson}},
  \bibinfo {author} {\bibfnamefont {K.~Jarrod}\ \bibnamefont {Millman}},
  \bibinfo {author} {\bibfnamefont {Nikolay}\ \bibnamefont {Mayorov}}, \bibinfo
  {author} {\bibfnamefont {Andrew R.~J.}\ \bibnamefont {Nelson}}, \bibinfo
  {author} {\bibfnamefont {Eric}\ \bibnamefont {Jones}}, \bibinfo {author}
  {\bibfnamefont {Robert}\ \bibnamefont {Kern}}, \bibinfo {author}
  {\bibfnamefont {Eric}\ \bibnamefont {Larson}}, \bibinfo {author}
  {\bibfnamefont {C~J}\ \bibnamefont {Carey}}, \bibinfo {author} {\bibfnamefont
  {{\.I}lhan}\ \bibnamefont {Polat}}, \bibinfo {author} {\bibfnamefont
  {Yu}~\bibnamefont {Feng}}, \bibinfo {author} {\bibfnamefont {Eric~W.}\
  \bibnamefont {Moore}}, \bibinfo {author} {\bibfnamefont {Jake}\ \bibnamefont
  {{VanderPlas}}}, \bibinfo {author} {\bibfnamefont {Denis}\ \bibnamefont
  {Laxalde}}, \bibinfo {author} {\bibfnamefont {Josef}\ \bibnamefont
  {Perktold}}, \bibinfo {author} {\bibfnamefont {Robert}\ \bibnamefont
  {Cimrman}}, \bibinfo {author} {\bibfnamefont {Ian}\ \bibnamefont
  {Henriksen}}, \bibinfo {author} {\bibfnamefont {E.~A.}\ \bibnamefont
  {Quintero}}, \bibinfo {author} {\bibfnamefont {Charles~R.}\ \bibnamefont
  {Harris}}, \bibinfo {author} {\bibfnamefont {Anne~M.}\ \bibnamefont
  {Archibald}}, \bibinfo {author} {\bibfnamefont {Ant{\^o}nio~H.}\ \bibnamefont
  {Ribeiro}}, \bibinfo {author} {\bibfnamefont {Fabian}\ \bibnamefont
  {Pedregosa}}, \bibinfo {author} {\bibfnamefont {Paul}\ \bibnamefont {{van
  Mulbregt}}}, \ and\ \bibinfo {author} {\bibnamefont {{SciPy 1.0
  Contributors}}},\ }\bibfield  {title} {\enquote {\bibinfo {title} {{{SciPy}
  1.0: Fundamental Algorithms for Scientific Computing in Python}},}\ }\href
  {\doibase 10.1038/s41592-019-0686-2} {\bibfield  {journal} {\bibinfo
  {journal} {Nature Methods}\ }\textbf {\bibinfo {volume} {17}},\ \bibinfo
  {pages} {261--272} (\bibinfo {year} {2020})}\BibitemShut {NoStop}%
\bibitem [{\citenamefont {Scheie}(2021)}]{Scheie2021py}%
  \BibitemOpen
  \bibfield  {author} {\bibinfo {author} {\bibfnamefont {A.}~\bibnamefont
  {Scheie}},\ }\bibfield  {title} {\enquote {\bibinfo {title} {{PyCrystalField:
  Software for Calculation, Analysis, and Fitting of Crystal Electric Field
  Hamiltonians}},}\ }\href {\doibase 10.1107/S160057672001554X} {\bibfield
  {journal} {\bibinfo  {journal} {J. Appl. Cryst.}\ }\textbf {\bibinfo {volume}
  {54}},\ \bibinfo {pages} {356} (\bibinfo {year} {2021})}\BibitemShut
  {NoStop}%
\bibitem [{\citenamefont {Xing}\ \emph {et~al.}(2021)\citenamefont {Xing},
  \citenamefont {Taddei}, \citenamefont {Sanjeewa}, \citenamefont {Fishman},
  \citenamefont {Daum}, \citenamefont {Mourigal}, \citenamefont {dela Cruz},\
  and\ \citenamefont {Sefat}}]{Xing2021}%
  \BibitemOpen
  \bibfield  {author} {\bibinfo {author} {\bibfnamefont {Jie}\ \bibnamefont
  {Xing}}, \bibinfo {author} {\bibfnamefont {Keith~M.}\ \bibnamefont {Taddei}},
  \bibinfo {author} {\bibfnamefont {Liurukara~D.}\ \bibnamefont {Sanjeewa}},
  \bibinfo {author} {\bibfnamefont {Randy~S.}\ \bibnamefont {Fishman}},
  \bibinfo {author} {\bibfnamefont {Marcus}\ \bibnamefont {Daum}}, \bibinfo
  {author} {\bibfnamefont {Martin}\ \bibnamefont {Mourigal}}, \bibinfo {author}
  {\bibfnamefont {C.}~\bibnamefont {dela Cruz}}, \ and\ \bibinfo {author}
  {\bibfnamefont {Athena~S.}\ \bibnamefont {Sefat}},\ }\bibfield  {title}
  {\enquote {\bibinfo {title} {Stripe antiferromagnetic ground state of the
  ideal triangular lattice compound ${\mathrm{kerse}}_{2}$},}\ }\href {\doibase
  10.1103/PhysRevB.103.144413} {\bibfield  {journal} {\bibinfo  {journal}
  {Phys. Rev. B}\ }\textbf {\bibinfo {volume} {103}},\ \bibinfo {pages}
  {144413} (\bibinfo {year} {2021})}\BibitemShut {NoStop}%
\bibitem [{\citenamefont {Cao}\ \emph {et~al.}(2016)\citenamefont {Cao},
  \citenamefont {Banerjee}, \citenamefont {Yan}, \citenamefont {Bridges},
  \citenamefont {Lumsden}, \citenamefont {Mandrus}, \citenamefont {Tennant},
  \citenamefont {Chakoumakos},\ and\ \citenamefont {Nagler}}]{HBCao2016}%
  \BibitemOpen
  \bibfield  {author} {\bibinfo {author} {\bibfnamefont {H.~B.}\ \bibnamefont
  {Cao}}, \bibinfo {author} {\bibfnamefont {A.}~\bibnamefont {Banerjee}},
  \bibinfo {author} {\bibfnamefont {J.-Q.}\ \bibnamefont {Yan}}, \bibinfo
  {author} {\bibfnamefont {C.~A.}\ \bibnamefont {Bridges}}, \bibinfo {author}
  {\bibfnamefont {M.~D.}\ \bibnamefont {Lumsden}}, \bibinfo {author}
  {\bibfnamefont {D.~G.}\ \bibnamefont {Mandrus}}, \bibinfo {author}
  {\bibfnamefont {D.~A.}\ \bibnamefont {Tennant}}, \bibinfo {author}
  {\bibfnamefont {B.~C.}\ \bibnamefont {Chakoumakos}}, \ and\ \bibinfo {author}
  {\bibfnamefont {S.~E.}\ \bibnamefont {Nagler}},\ }\bibfield  {title}
  {\enquote {\bibinfo {title} {Low-temperature crystal and magnetic structure
  of $\ensuremath{\alpha}\ensuremath{-}{\mathrm{rucl}}_{3}$},}\ }\href
  {\doibase 10.1103/PhysRevB.93.134423} {\bibfield  {journal} {\bibinfo
  {journal} {Phys. Rev. B}\ }\textbf {\bibinfo {volume} {93}},\ \bibinfo
  {pages} {134423} (\bibinfo {year} {2016})}\BibitemShut {NoStop}%
\bibitem [{\citenamefont {Bette}\ \emph {et~al.}(2019)\citenamefont {Bette},
  \citenamefont {Takayama}, \citenamefont {Duppel}, \citenamefont {Poulain},
  \citenamefont {Takagi},\ and\ \citenamefont {Dinnebier}}]{Bette2019}%
  \BibitemOpen
  \bibfield  {author} {\bibinfo {author} {\bibfnamefont {Sebastian}\
  \bibnamefont {Bette}}, \bibinfo {author} {\bibfnamefont {Tomohiro}\
  \bibnamefont {Takayama}}, \bibinfo {author} {\bibfnamefont {Viola}\
  \bibnamefont {Duppel}}, \bibinfo {author} {\bibfnamefont {Agnieszka}\
  \bibnamefont {Poulain}}, \bibinfo {author} {\bibfnamefont {Hidenori}\
  \bibnamefont {Takagi}}, \ and\ \bibinfo {author} {\bibfnamefont {Robert~E.}\
  \bibnamefont {Dinnebier}},\ }\bibfield  {title} {\enquote {\bibinfo {title}
  {Crystal structure and stacking faults in the layered honeycomb{,}
  delafossite-type materials ag3liir2o6 and ag3liru2o6},}\ }\href {\doibase
  10.1039/C9DT01789E} {\bibfield  {journal} {\bibinfo  {journal} {Dalton
  Trans.}\ }\textbf {\bibinfo {volume} {48}},\ \bibinfo {pages} {9250--9259}
  (\bibinfo {year} {2019})}\BibitemShut {NoStop}%
\bibitem [{\citenamefont {Viciu}\ \emph {et~al.}(2006)\citenamefont {Viciu},
  \citenamefont {Bos}, \citenamefont {Zandbergen}, \citenamefont {Huang},
  \citenamefont {Foo}, \citenamefont {Ishiwata}, \citenamefont {Ramirez},
  \citenamefont {Lee}, \citenamefont {Ong},\ and\ \citenamefont {Cava}}]{nco}%
  \BibitemOpen
  \bibfield  {author} {\bibinfo {author} {\bibfnamefont {L.}~\bibnamefont
  {Viciu}}, \bibinfo {author} {\bibfnamefont {J.~W.~G.}\ \bibnamefont {Bos}},
  \bibinfo {author} {\bibfnamefont {H.~W.}\ \bibnamefont {Zandbergen}},
  \bibinfo {author} {\bibfnamefont {Q.}~\bibnamefont {Huang}}, \bibinfo
  {author} {\bibfnamefont {M.~L.}\ \bibnamefont {Foo}}, \bibinfo {author}
  {\bibfnamefont {S.}~\bibnamefont {Ishiwata}}, \bibinfo {author}
  {\bibfnamefont {A.~P.}\ \bibnamefont {Ramirez}}, \bibinfo {author}
  {\bibfnamefont {M.}~\bibnamefont {Lee}}, \bibinfo {author} {\bibfnamefont
  {N.~P.}\ \bibnamefont {Ong}}, \ and\ \bibinfo {author} {\bibfnamefont
  {R.~J.}\ \bibnamefont {Cava}},\ }\bibfield  {title} {\enquote {\bibinfo
  {title} {Crystal structure and elementary properties of
  ${\mathrm{na}}_{x}\mathrm{Co}{\mathrm{o}}_{2}$ ($x=0.32$, 0.51, 0.6, 0.75,
  and 0.92) in the three-layer $\mathrm{Na}\mathrm{Co}{\mathrm{o}}_{2}$
  family},}\ }\href {\doibase 10.1103/PhysRevB.73.174104} {\bibfield  {journal}
  {\bibinfo  {journal} {Phys. Rev. B}\ }\textbf {\bibinfo {volume} {73}},\
  \bibinfo {pages} {174104} (\bibinfo {year} {2006})}\BibitemShut {NoStop}%
\bibitem [{\citenamefont {Blundell}(2001)}]{Blundellbook}%
  \BibitemOpen
  \bibfield  {author} {\bibinfo {author} {\bibfnamefont {Stephen}\ \bibnamefont
  {Blundell}},\ }\href@noop {} {\emph {\bibinfo {title} {Magnetism in Condensed
  Matter}}}\ (\bibinfo  {publisher} {Oxford University Press},\ \bibinfo
  {address} {Oxford, UK},\ \bibinfo {year} {2001})\BibitemShut {NoStop}%
\bibitem [{sup()}]{supp}%
  \BibitemOpen
  \href@noop {} {}\bibinfo {note} {See the Supplementary
  Materials.}\BibitemShut {Stop}%
\bibitem [{\citenamefont {Rotundu}\ \emph {et~al.}(2004)\citenamefont
  {Rotundu}, \citenamefont {Tsujii}, \citenamefont {Takano}, \citenamefont
  {Andraka}, \citenamefont {Sugawara}, \citenamefont {Aoki},\ and\
  \citenamefont {Sato}}]{Rotundu2004}%
  \BibitemOpen
  \bibfield  {author} {\bibinfo {author} {\bibfnamefont {C.~R.}\ \bibnamefont
  {Rotundu}}, \bibinfo {author} {\bibfnamefont {H.}~\bibnamefont {Tsujii}},
  \bibinfo {author} {\bibfnamefont {Y.}~\bibnamefont {Takano}}, \bibinfo
  {author} {\bibfnamefont {B.}~\bibnamefont {Andraka}}, \bibinfo {author}
  {\bibfnamefont {H.}~\bibnamefont {Sugawara}}, \bibinfo {author}
  {\bibfnamefont {Y.}~\bibnamefont {Aoki}}, \ and\ \bibinfo {author}
  {\bibfnamefont {H.}~\bibnamefont {Sato}},\ }\bibfield  {title} {\enquote
  {\bibinfo {title} {High magnetic field phase diagram of
  ${\mathrm{p}\mathrm{r}\mathrm{o}\mathrm{s}}_{4}{\mathrm{s}\mathrm{b}}_{12}$},}\
  }\href {\doibase 10.1103/PhysRevLett.92.037203} {\bibfield  {journal}
  {\bibinfo  {journal} {Phys. Rev. Lett.}\ }\textbf {\bibinfo {volume} {92}},\
  \bibinfo {pages} {037203} (\bibinfo {year} {2004})}\BibitemShut {NoStop}%
\bibitem [{\citenamefont {Bordelon}\ \emph {et~al.}(2020)\citenamefont
  {Bordelon}, \citenamefont {Liu}, \citenamefont {Posthuma}, \citenamefont
  {Sarte}, \citenamefont {Butch}, \citenamefont {Pajerowski}, \citenamefont
  {Banerjee}, \citenamefont {Balents},\ and\ \citenamefont
  {Wilson}}]{Bordelon2020}%
  \BibitemOpen
  \bibfield  {author} {\bibinfo {author} {\bibfnamefont {Mitchell~M.}\
  \bibnamefont {Bordelon}}, \bibinfo {author} {\bibfnamefont {Chunxiao}\
  \bibnamefont {Liu}}, \bibinfo {author} {\bibfnamefont {Lorenzo}\ \bibnamefont
  {Posthuma}}, \bibinfo {author} {\bibfnamefont {P.~M.}\ \bibnamefont {Sarte}},
  \bibinfo {author} {\bibfnamefont {N.~P.}\ \bibnamefont {Butch}}, \bibinfo
  {author} {\bibfnamefont {Daniel~M.}\ \bibnamefont {Pajerowski}}, \bibinfo
  {author} {\bibfnamefont {Arnab}\ \bibnamefont {Banerjee}}, \bibinfo {author}
  {\bibfnamefont {Leon}\ \bibnamefont {Balents}}, \ and\ \bibinfo {author}
  {\bibfnamefont {Stephen~D.}\ \bibnamefont {Wilson}},\ }\bibfield  {title}
  {\enquote {\bibinfo {title} {{Spin excitations in the frustrated triangular
  lattice antiferromagnet ${\mathrm{NaYbO}}_{2}$}},}\ }\href {\doibase
  10.1103/PhysRevB.101.224427} {\bibfield  {journal} {\bibinfo  {journal}
  {Phys. Rev. B}\ }\textbf {\bibinfo {volume} {101}},\ \bibinfo {pages}
  {224427} (\bibinfo {year} {2020})}\BibitemShut {NoStop}%
\bibitem [{\citenamefont {Dun}\ \emph {et~al.}(2021)\citenamefont {Dun},
  \citenamefont {Bai}, \citenamefont {Stone}, \citenamefont {Zhou},\ and\
  \citenamefont {Mourigal}}]{Dun2021}%
  \BibitemOpen
  \bibfield  {author} {\bibinfo {author} {\bibfnamefont {Zhiling}\ \bibnamefont
  {Dun}}, \bibinfo {author} {\bibfnamefont {Xiaojian}\ \bibnamefont {Bai}},
  \bibinfo {author} {\bibfnamefont {Matthew~B.}\ \bibnamefont {Stone}},
  \bibinfo {author} {\bibfnamefont {Haidong}\ \bibnamefont {Zhou}}, \ and\
  \bibinfo {author} {\bibfnamefont {Martin}\ \bibnamefont {Mourigal}},\
  }\bibfield  {title} {\enquote {\bibinfo {title} {Effective point-charge
  analysis of crystal fields: Application to rare-earth pyrochlores and tripod
  kagome magnets
  $r{}_{3}\mathrm{Mg}{}_{2}\mathrm{Sb}{}_{3}\mathrm{O}{}_{14}$},}\ }\href
  {\doibase 10.1103/PhysRevResearch.3.023012} {\bibfield  {journal} {\bibinfo
  {journal} {Phys. Rev. Res.}\ }\textbf {\bibinfo {volume} {3}},\ \bibinfo
  {pages} {023012} (\bibinfo {year} {2021})}\BibitemShut {NoStop}%
\bibitem [{\citenamefont {Scheie}(2022)}]{Scheie_scipost2022}%
  \BibitemOpen
  \bibfield  {author} {\bibinfo {author} {\bibfnamefont {Allen}\ \bibnamefont
  {Scheie}},\ }\bibfield  {title} {\enquote {\bibinfo {title} {{Quantifying
  uncertainties in crystal electric field Hamiltonian fits to neutron data}},}\
  }\href {\doibase 10.21468/SciPostPhysCore.5.1.018} {\bibfield  {journal}
  {\bibinfo  {journal} {SciPost Phys. Core}\ }\textbf {\bibinfo {volume} {5}},\
  \bibinfo {pages} {018} (\bibinfo {year} {2022})}\BibitemShut {NoStop}%
\bibitem [{\citenamefont {Taherunnisa}\ \emph {et~al.}(2019)\citenamefont
  {Taherunnisa}, \citenamefont {Reddy}, \citenamefont {Rao}, \citenamefont
  {Rudramamba}, \citenamefont {Zhydachevskyy}, \citenamefont {Suchocki},
  \citenamefont {Piasecki},\ and\ \citenamefont {Reddy}}]{Taherunnisa2019}%
  \BibitemOpen
  \bibfield  {author} {\bibinfo {author} {\bibfnamefont {SK}~\bibnamefont
  {Taherunnisa}}, \bibinfo {author} {\bibfnamefont {D.V.~Krishna}\ \bibnamefont
  {Reddy}}, \bibinfo {author} {\bibfnamefont {T.~Sambasiva}\ \bibnamefont
  {Rao}}, \bibinfo {author} {\bibfnamefont {K.S.}\ \bibnamefont {Rudramamba}},
  \bibinfo {author} {\bibfnamefont {Y.A.}\ \bibnamefont {Zhydachevskyy}},
  \bibinfo {author} {\bibfnamefont {A.}~\bibnamefont {Suchocki}}, \bibinfo
  {author} {\bibfnamefont {M.}~\bibnamefont {Piasecki}}, \ and\ \bibinfo
  {author} {\bibfnamefont {M.~Rami}\ \bibnamefont {Reddy}},\ }\bibfield
  {title} {\enquote {\bibinfo {title} {Effect of up-conversion luminescence in
  er$^{3+}$ doped phosphate glasses for developing erbium-doped fibre
  amplifiers (edfa) and g-led's},}\ }\href {\doibase
  https://doi.org/10.1016/j.omx.2019.100034} {\bibfield  {journal} {\bibinfo
  {journal} {Optical Materials: X}\ }\textbf {\bibinfo {volume} {3}},\ \bibinfo
  {pages} {100034} (\bibinfo {year} {2019})}\BibitemShut {NoStop}%
\bibitem [{\citenamefont {Becker}\ \emph {et~al.}(2025)\citenamefont {Becker},
  \citenamefont {Curtin}, \citenamefont {KC}, \citenamefont {Schneider},
  \citenamefont {Sauerzopf}, \citenamefont {Elzeiny},\ and\ \citenamefont
  {M\"uller}}]{Becker2025}%
  \BibitemOpen
  \bibfield  {author} {\bibinfo {author} {\bibfnamefont {Fabian}\ \bibnamefont
  {Becker}}, \bibinfo {author} {\bibfnamefont {Catherine~L.}\ \bibnamefont
  {Curtin}}, \bibinfo {author} {\bibfnamefont {Sudip}\ \bibnamefont {KC}},
  \bibinfo {author} {\bibfnamefont {Tim}\ \bibnamefont {Schneider}}, \bibinfo
  {author} {\bibfnamefont {Lorenz J.~J.}\ \bibnamefont {Sauerzopf}}, \bibinfo
  {author} {\bibfnamefont {Ibrahim}\ \bibnamefont {Elzeiny}}, \ and\ \bibinfo
  {author} {\bibfnamefont {Kai}\ \bibnamefont {M\"uller}},\ }\bibfield  {title}
  {\enquote {\bibinfo {title} {Spectroscopic investigations of multiple
  environments in er:${\mathrm{cawo}}_{4}$ through charge imbalance},}\ }\href
  {\doibase 10.1103/6srk-3k4n} {\bibfield  {journal} {\bibinfo  {journal}
  {Phys. Rev. Mater.}\ }\textbf {\bibinfo {volume} {9}},\ \bibinfo {pages}
  {076203} (\bibinfo {year} {2025})}\BibitemShut {NoStop}%
\bibitem [{\citenamefont {Ozerov}\ \emph {et~al.}(2022)\citenamefont {Ozerov},
  \citenamefont {Anand}, \citenamefont {van~de Burgt}, \citenamefont {Lu},
  \citenamefont {Holleman}, \citenamefont {Zhou}, \citenamefont {McGill},\ and\
  \citenamefont {Beekman}}]{Ozerov2022}%
  \BibitemOpen
  \bibfield  {author} {\bibinfo {author} {\bibfnamefont {Mykhaylo}\
  \bibnamefont {Ozerov}}, \bibinfo {author} {\bibfnamefont {Naween}\
  \bibnamefont {Anand}}, \bibinfo {author} {\bibfnamefont {L.~J.}\ \bibnamefont
  {van~de Burgt}}, \bibinfo {author} {\bibfnamefont {Zhengguang}\ \bibnamefont
  {Lu}}, \bibinfo {author} {\bibfnamefont {Jade}\ \bibnamefont {Holleman}},
  \bibinfo {author} {\bibfnamefont {Haidong}\ \bibnamefont {Zhou}}, \bibinfo
  {author} {\bibfnamefont {Steve}\ \bibnamefont {McGill}}, \ and\ \bibinfo
  {author} {\bibfnamefont {Christianne}\ \bibnamefont {Beekman}},\ }\bibfield
  {title} {\enquote {\bibinfo {title} {Magnetic field tuning of crystal field
  levels and vibronic states in the spin ice compound
  ${\mathrm{ho}}_{2}{\mathrm{ti}}_{2}{\mathrm{o}}_{7}$ observed with far
  infrared reflectometry},}\ }\href {\doibase 10.1103/PhysRevB.105.165102}
  {\bibfield  {journal} {\bibinfo  {journal} {Phys. Rev. B}\ }\textbf {\bibinfo
  {volume} {105}},\ \bibinfo {pages} {165102} (\bibinfo {year}
  {2022})}\BibitemShut {NoStop}%
\bibitem [{\citenamefont {Xiang}\ \emph {et~al.}(2023)\citenamefont {Xiang},
  \citenamefont {Dhakal}, \citenamefont {Ozerov}, \citenamefont {Jiang},
  \citenamefont {Mou}, \citenamefont {Ozarowski}, \citenamefont {Huang},
  \citenamefont {Zhou}, \citenamefont {Fang}, \citenamefont {Winter},
  \citenamefont {Jiang},\ and\ \citenamefont {Smirnov}}]{LXiang2023}%
  \BibitemOpen
  \bibfield  {author} {\bibinfo {author} {\bibfnamefont {Li}~\bibnamefont
  {Xiang}}, \bibinfo {author} {\bibfnamefont {Ramesh}\ \bibnamefont {Dhakal}},
  \bibinfo {author} {\bibfnamefont {Mykhaylo}\ \bibnamefont {Ozerov}}, \bibinfo
  {author} {\bibfnamefont {Yuxuan}\ \bibnamefont {Jiang}}, \bibinfo {author}
  {\bibfnamefont {Banasree~S.}\ \bibnamefont {Mou}}, \bibinfo {author}
  {\bibfnamefont {Andrew}\ \bibnamefont {Ozarowski}}, \bibinfo {author}
  {\bibfnamefont {Qing}\ \bibnamefont {Huang}}, \bibinfo {author}
  {\bibfnamefont {Haidong}\ \bibnamefont {Zhou}}, \bibinfo {author}
  {\bibfnamefont {Jiyuan}\ \bibnamefont {Fang}}, \bibinfo {author}
  {\bibfnamefont {Stephen~M.}\ \bibnamefont {Winter}}, \bibinfo {author}
  {\bibfnamefont {Zhigang}\ \bibnamefont {Jiang}}, \ and\ \bibinfo {author}
  {\bibfnamefont {Dmitry}\ \bibnamefont {Smirnov}},\ }\bibfield  {title}
  {\enquote {\bibinfo {title} {Disorder-enriched magnetic excitations in a
  heisenberg-kitaev quantum magnet
  ${\mathrm{na}}_{2}{\mathrm{co}}_{2}{\mathrm{teo}}_{6}$},}\ }\href {\doibase
  10.1103/PhysRevLett.131.076701} {\bibfield  {journal} {\bibinfo  {journal}
  {Phys. Rev. Lett.}\ }\textbf {\bibinfo {volume} {131}},\ \bibinfo {pages}
  {076701} (\bibinfo {year} {2023})}\BibitemShut {NoStop}%
\bibitem [{\citenamefont {Mou}\ \emph {et~al.}(2024)\citenamefont {Mou},
  \citenamefont {Zhang}, \citenamefont {Xiang}, \citenamefont {Xu},
  \citenamefont {Zhong}, \citenamefont {Cava}, \citenamefont {Zhou},
  \citenamefont {Jiang}, \citenamefont {Smirnov}, \citenamefont {Drichko},\
  and\ \citenamefont {Winter}}]{Mou2024}%
  \BibitemOpen
  \bibfield  {author} {\bibinfo {author} {\bibfnamefont {Banasree~S.}\
  \bibnamefont {Mou}}, \bibinfo {author} {\bibfnamefont {Xinshu}\ \bibnamefont
  {Zhang}}, \bibinfo {author} {\bibfnamefont {Li}~\bibnamefont {Xiang}},
  \bibinfo {author} {\bibfnamefont {Yuanyuan}\ \bibnamefont {Xu}}, \bibinfo
  {author} {\bibfnamefont {Ruidan}\ \bibnamefont {Zhong}}, \bibinfo {author}
  {\bibfnamefont {Robert~J.}\ \bibnamefont {Cava}}, \bibinfo {author}
  {\bibfnamefont {Haidong}\ \bibnamefont {Zhou}}, \bibinfo {author}
  {\bibfnamefont {Zhigang}\ \bibnamefont {Jiang}}, \bibinfo {author}
  {\bibfnamefont {Dmitry}\ \bibnamefont {Smirnov}}, \bibinfo {author}
  {\bibfnamefont {Natalia}\ \bibnamefont {Drichko}}, \ and\ \bibinfo {author}
  {\bibfnamefont {Stephen~M.}\ \bibnamefont {Winter}},\ }\bibfield  {title}
  {\enquote {\bibinfo {title} {Comparative raman scattering study of crystal
  field excitations in co-based quantum magnets},}\ }\href {\doibase
  10.1103/PhysRevMaterials.8.084408} {\bibfield  {journal} {\bibinfo  {journal}
  {Phys. Rev. Mater.}\ }\textbf {\bibinfo {volume} {8}},\ \bibinfo {pages}
  {084408} (\bibinfo {year} {2024})}\BibitemShut {NoStop}%
\bibitem [{\citenamefont {Lujan}\ \emph {et~al.}(2024)\citenamefont {Lujan},
  \citenamefont {Choe}, \citenamefont {Chaudhary}, \citenamefont {Ye},
  \citenamefont {Nnokwe}, \citenamefont {Rodriguez-Vega}, \citenamefont {He},
  \citenamefont {Gao}, \citenamefont {Nunley}, \citenamefont {Baldini},
  \citenamefont {Zhou}, \citenamefont {Fiete}, \citenamefont {He},\ and\
  \citenamefont {Li}}]{Lujan2024}%
  \BibitemOpen
  \bibfield  {author} {\bibinfo {author} {\bibfnamefont {David}\ \bibnamefont
  {Lujan}}, \bibinfo {author} {\bibfnamefont {Jeongheon}\ \bibnamefont {Choe}},
  \bibinfo {author} {\bibfnamefont {Swati}\ \bibnamefont {Chaudhary}}, \bibinfo
  {author} {\bibfnamefont {Gaihua}\ \bibnamefont {Ye}}, \bibinfo {author}
  {\bibfnamefont {Cynthia}\ \bibnamefont {Nnokwe}}, \bibinfo {author}
  {\bibfnamefont {Martin}\ \bibnamefont {Rodriguez-Vega}}, \bibinfo {author}
  {\bibfnamefont {Jiaming}\ \bibnamefont {He}}, \bibinfo {author}
  {\bibfnamefont {Frank~Y.}\ \bibnamefont {Gao}}, \bibinfo {author}
  {\bibfnamefont {T.~Nathan}\ \bibnamefont {Nunley}}, \bibinfo {author}
  {\bibfnamefont {Edoardo}\ \bibnamefont {Baldini}}, \bibinfo {author}
  {\bibfnamefont {Jianshi}\ \bibnamefont {Zhou}}, \bibinfo {author}
  {\bibfnamefont {Gregory~A.}\ \bibnamefont {Fiete}}, \bibinfo {author}
  {\bibfnamefont {Rui}\ \bibnamefont {He}}, \ and\ \bibinfo {author}
  {\bibfnamefont {Xiaoqin}\ \bibnamefont {Li}},\ }\bibfield  {title} {\enquote
  {\bibinfo {title} {Spin–orbit exciton–induced phonon chirality in a
  quantum magnet},}\ }\href {\doibase 10.1073/pnas.2304360121} {\bibfield
  {journal} {\bibinfo  {journal} {Proceedings of the National Academy of
  Sciences}\ }\textbf {\bibinfo {volume} {121}},\ \bibinfo {pages}
  {e2304360121} (\bibinfo {year} {2024})}\BibitemShut {NoStop}%
\bibitem [{\citenamefont {Mai}\ \emph {et~al.}(2025)\citenamefont {Mai},
  \citenamefont {Li}, \citenamefont {Garrity}, \citenamefont {Shaw},
  \citenamefont {DeLazzer}, \citenamefont {Dally}, \citenamefont {Adel},
  \citenamefont {Mu\~noz}, \citenamefont {Giovannone}, \citenamefont {Lyon},
  \citenamefont {Pawbake}, \citenamefont {Faugeras}, \citenamefont
  {Le~Mardele}, \citenamefont {Orlita}, \citenamefont {Simpson}, \citenamefont
  {Ross}, \citenamefont {Aguilar},\ and\ \citenamefont {Walker}}]{Mai2025}%
  \BibitemOpen
  \bibfield  {author} {\bibinfo {author} {\bibfnamefont {Thuc~T.}\ \bibnamefont
  {Mai}}, \bibinfo {author} {\bibfnamefont {Yufei}\ \bibnamefont {Li}},
  \bibinfo {author} {\bibfnamefont {K.~F.}\ \bibnamefont {Garrity}}, \bibinfo
  {author} {\bibfnamefont {D.}~\bibnamefont {Shaw}}, \bibinfo {author}
  {\bibfnamefont {T.}~\bibnamefont {DeLazzer}}, \bibinfo {author}
  {\bibfnamefont {R.~L.}\ \bibnamefont {Dally}}, \bibinfo {author}
  {\bibfnamefont {T.}~\bibnamefont {Adel}}, \bibinfo {author} {\bibfnamefont
  {M.~F.}\ \bibnamefont {Mu\~noz}}, \bibinfo {author} {\bibfnamefont
  {A.}~\bibnamefont {Giovannone}}, \bibinfo {author} {\bibfnamefont
  {C.}~\bibnamefont {Lyon}}, \bibinfo {author} {\bibfnamefont {A.}~\bibnamefont
  {Pawbake}}, \bibinfo {author} {\bibfnamefont {C.}~\bibnamefont {Faugeras}},
  \bibinfo {author} {\bibfnamefont {F.}~\bibnamefont {Le~Mardele}}, \bibinfo
  {author} {\bibfnamefont {M.}~\bibnamefont {Orlita}}, \bibinfo {author}
  {\bibfnamefont {J.~R.}\ \bibnamefont {Simpson}}, \bibinfo {author}
  {\bibfnamefont {K.}~\bibnamefont {Ross}}, \bibinfo {author} {\bibfnamefont
  {R.~Vald\'es}\ \bibnamefont {Aguilar}}, \ and\ \bibinfo {author}
  {\bibfnamefont {A.~R.~Hight}\ \bibnamefont {Walker}},\ }\bibfield  {title}
  {\enquote {\bibinfo {title} {Spin-orbital--lattice coupling and the phonon
  zeeman effect in the dirac honeycomb magnet ${\mathrm{cotio}}_{3}$},}\ }\href
  {\doibase 10.1103/PhysRevB.111.104419} {\bibfield  {journal} {\bibinfo
  {journal} {Phys. Rev. B}\ }\textbf {\bibinfo {volume} {111}},\ \bibinfo
  {pages} {104419} (\bibinfo {year} {2025})}\BibitemShut {NoStop}%
\bibitem [{\citenamefont {Liu}\ \emph {et~al.}(2018)\citenamefont {Liu},
  \citenamefont {\ifmmode~\check{C}\else \v{C}\fi{}erm\'ak}, \citenamefont
  {Franz}, \citenamefont {Pfleiderer},\ and\ \citenamefont
  {Schneidewind}}]{BQLiu2018}%
  \BibitemOpen
  \bibfield  {author} {\bibinfo {author} {\bibfnamefont {B.-Q.}\ \bibnamefont
  {Liu}}, \bibinfo {author} {\bibfnamefont {P.}~\bibnamefont
  {\ifmmode~\check{C}\else \v{C}\fi{}erm\'ak}}, \bibinfo {author}
  {\bibfnamefont {C.}~\bibnamefont {Franz}}, \bibinfo {author} {\bibfnamefont
  {C.}~\bibnamefont {Pfleiderer}}, \ and\ \bibinfo {author} {\bibfnamefont
  {A.}~\bibnamefont {Schneidewind}},\ }\bibfield  {title} {\enquote {\bibinfo
  {title} {{Lattice dynamics and coupled quadrupole-phonon excitations in
  ${\mathrm{CeAuAl}}_{3}$}},}\ }\href {\doibase 10.1103/PhysRevB.98.174306}
  {\bibfield  {journal} {\bibinfo  {journal} {Phys. Rev. B}\ }\textbf {\bibinfo
  {volume} {98}},\ \bibinfo {pages} {174306} (\bibinfo {year}
  {2018})}\BibitemShut {NoStop}%
\bibitem [{\citenamefont {{\v C}erm{\'a}k}\ \emph {et~al.}(2019)\citenamefont
  {{\v C}erm{\'a}k}, \citenamefont {Schneidewind}, \citenamefont {Liu},
  \citenamefont {Koza}, \citenamefont {Franz}, \citenamefont {Sch{\"o}nmann},
  \citenamefont {Sobolev},\ and\ \citenamefont {Pfleiderer}}]{Cermak2019}%
  \BibitemOpen
  \bibfield  {author} {\bibinfo {author} {\bibfnamefont {Petr}\ \bibnamefont
  {{\v C}erm{\'a}k}}, \bibinfo {author} {\bibfnamefont {Astrid}\ \bibnamefont
  {Schneidewind}}, \bibinfo {author} {\bibfnamefont {Benqiong}\ \bibnamefont
  {Liu}}, \bibinfo {author} {\bibfnamefont {Michael~Marek}\ \bibnamefont
  {Koza}}, \bibinfo {author} {\bibfnamefont {Christian}\ \bibnamefont {Franz}},
  \bibinfo {author} {\bibfnamefont {Rudolf}\ \bibnamefont {Sch{\"o}nmann}},
  \bibinfo {author} {\bibfnamefont {Oleg}\ \bibnamefont {Sobolev}}, \ and\
  \bibinfo {author} {\bibfnamefont {Christian}\ \bibnamefont {Pfleiderer}},\
  }\bibfield  {title} {\enquote {\bibinfo {title} {{Magnetoelastic hybrid
  excitations in CeAuAl$_3$}},}\ }\href {\doibase 10.1073/pnas.1819664116}
  {\bibfield  {journal} {\bibinfo  {journal} {Proc. Natl. Acad. Sci.}\ }\textbf
  {\bibinfo {volume} {116}},\ \bibinfo {pages} {6695} (\bibinfo {year}
  {2019})}\BibitemShut {NoStop}%
\bibitem [{\citenamefont {Pocs}\ \emph {et~al.}(2025)\citenamefont {Pocs},
  \citenamefont {Leahy}, \citenamefont {Xing}, \citenamefont {Choi},
  \citenamefont {Sefat}, \citenamefont {Hermele},\ and\ \citenamefont
  {Lee}}]{Pocs2025}%
  \BibitemOpen
  \bibfield  {author} {\bibinfo {author} {\bibfnamefont {Christopher~A.}\
  \bibnamefont {Pocs}}, \bibinfo {author} {\bibfnamefont {Ian~A.}\ \bibnamefont
  {Leahy}}, \bibinfo {author} {\bibfnamefont {Jie}\ \bibnamefont {Xing}},
  \bibinfo {author} {\bibfnamefont {Eun~Sang}\ \bibnamefont {Choi}}, \bibinfo
  {author} {\bibfnamefont {Athena~S.}\ \bibnamefont {Sefat}}, \bibinfo {author}
  {\bibfnamefont {Michael}\ \bibnamefont {Hermele}}, \ and\ \bibinfo {author}
  {\bibfnamefont {Minhyea}\ \bibnamefont {Lee}},\ }\bibfield  {title} {\enquote
  {\bibinfo {title} {Heat conduction in magnetic insulators via hybridization
  of acoustic phonons and spin-flip excitations},}\ }\href {\doibase
  10.1103/PhysRevResearch.7.L022007} {\bibfield  {journal} {\bibinfo  {journal}
  {Phys. Rev. Res.}\ }\textbf {\bibinfo {volume} {7}},\ \bibinfo {pages}
  {L022007} (\bibinfo {year} {2025})}\BibitemShut {NoStop}%
\end{thebibliography}
%

\end{document}